\documentclass[twocolumn,english,aps,floatfix,superscriptaddress,amsfonts,amssymb,superscriptaddress]{revtex4}
\usepackage{amsmath}
\usepackage{graphicx}
\usepackage{amssymb}
\usepackage{dsfont}
\usepackage{color}
\usepackage{lipsum,kantlipsum}
\usepackage[bookmarks=true,colorlinks,linkcolor=blue,urlcolor=blue,citecolor=blue]{hyperref}
\usepackage{mathtools}
\usepackage{amsthm}
\theoremstyle{plain}

\newcommand{\beq}{\begin{equation}}
\newcommand{\eeq}{\end{equation}}
\newcommand{\bea}{\begin{eqnarray}}
\newcommand{\eea}{\end{eqnarray}}

\newcommand{\eq}{\begin{equation}}
\newcommand{\en}{\end{equation}}
\newcommand{\ear}{\begin{eqnarray}}
\newcommand{\rae}{\end{eqnarray}}

\newcommand{\be}{\begin{eqnarray}}
\newcommand{\ee}{\end{eqnarray}}

\newcommand{\non}{\nonumber}

\usepackage[most]{tcolorbox}

\newtheorem*{algo*}{Algorithm}

\numberwithin{equation}{section}

\makeindex

\begin{document}
\title{Ising analogues of quantum spin chains with multispin interactions
}
\author{Francisco C. Alcaraz}
\email{alcaraz@ifsc.usp.br}
\affiliation{ Instituto de F\'{\i}sica de S\~{a}o Carlos, Universidade de S\~{a}o Paulo,
Caixa Postal 369, 13560-970, S\~{a}o Carlos, SP, Brazil}

\author{Rodrigo A. Pimenta}
\email{rodrigo.alvespimenta@umanitoba.ca}
\affiliation{Department of Physics and Astronomy, University of Manitoba, Winnipeg R3T 2N2, Canada}

\author{Jesko Sirker}
\email{sirker@physics.umanitoba.ca}
\affiliation{Department of Physics and Astronomy, University of Manitoba, Winnipeg R3T 2N2, Canada}
\affiliation{Manitoba Quantum Institute, University of Manitoba, Winnipeg R3T 2N2, Canada}

\date{\today{}}

\begin{abstract}
A new family of free fermionic quantum spin chains with multispin interactions was recently introduced. Here we show that it is possible to build standard quantum Ising chains---but with inhomogeneous couplings---which have the same spectra as the novel spin chains with multispin interactions. 
The Ising models are obtained by associating an antisymmetric tridiagonal matrix to the polynomials that characterize the quasienergies of the system via a modified Euclidean algorithm.
For the simplest non-trivial case, corresponding to the Fendley model, the phase diagram of the inhomogeneous Ising model is investigated numerically. It is characterized by gapped phases separated by critical lines with order-disorder transitions depending on the parity of the total number of energy density operators in the Hamiltonian.
\end{abstract}

\maketitle

\section{Introduction}\label{sec:intro}

Some of the simplest models of quantum matter are quantum spin chains which can be mapped onto Gaussian fermionic models. They are amenable to exact solutions and can provide insights into many-body phenomena. A quantum system has a free fermionic spectrum
if its Hamiltonian can be expressed in a quadratic form in 
terms of  (Dirac or Majorana) fermionic operators. The most common framework to solve free fermionic systems dates back to the seminal
work of Onsager \cite{O44}, later simplified by Kaufman \cite{K64}
and Lieb, Schultz and Mattis \cite{SML64}.

One example of a free-fermionic Hamiltonian is the inhomogeneous
 quantum Ising chain in a transverse field with open boundary conditions (OBC) \cite{P70},
\be\label{Ising}
H=-\sum_{\ell=1}^{L-1}w_{2\ell}\sigma_\ell^z\sigma_{\ell+1}^z-\sum_{\ell=1}^Lw_{2\ell-1}\sigma_\ell^x\,,
\ee
where $w_\ell$ are nonnegative real coupling parameters and the standard Pauli
matrices $\sigma_i^{x,z}$ act on the $i$-th site of the chain with $L$ sites. To uncover the free fermionic structure,
one maps the Pauli matrices
either to complex fermionic operators $c_\ell$ or to Majorana operators $\psi_{\ell}$ through the Jordan-Wigner transformation
\begin{eqnarray}
\label{JW}
\sigma_\ell^x &=& 1-2c_\ell^\dagger c_\ell = i\psi_{2\ell-1}\psi_{2\ell} \nonumber \\
\sigma_\ell^z &=& -\prod_{m=1}^{\ell-1}(1-2c_m^\dagger c_m)(c_\ell+c_\ell^\dagger) \\
&=& \prod_{m=1}^{\ell-1}i\psi_{2m-1}\psi_{2m}\psi_{2\ell-1} \nonumber
\end{eqnarray}
leading to
\begin{eqnarray}
\label{quadratic}
H&=&-\sum_{\ell=1}^{L-1} w_{2\ell}(c_\ell^\dagger c_{\ell+1}+c_\ell^\dagger c_{\ell+1}^\dagger+h.c.) \nonumber \\
&& -\sum_{\ell=1}^L w_{2\ell-1}(1-2c_\ell^\dagger c_\ell) \\
&=& i\sum_{\ell=1}^{2L-1}w_{\ell}\psi_{\ell}\psi_{\ell+1}. \nonumber
\end{eqnarray}
The operators $\psi_\ell$ satisfy the Clifford algebra
\be\label{Clifford}
&&\{\psi_\ell,\psi_{\ell'}\} = 2\delta_{\ell\ell'}
\ee
whereas the operators $c_\ell$ fulfill the complex fermionic algebra
\be\label{fermions}
&&\{c_\ell, c_{\ell'}\} = \{c_\ell^\dagger, c_{\ell'}^\dagger\}=0,\quad \{c_\ell^\dagger, c_{\ell'}\} =
\delta_{\ell\ell'}\,.
\ee
The two representations are connected by
\begin{equation}
    \label{ctopsi}
    c_\ell=\frac{1}{2}(\psi_{2\ell-1}-\text{i}\psi_{2\ell}),\quad c_\ell^\dagger=\frac{1}{2}(\psi_{2\ell-1}+\text{i}\psi_{2\ell})\, .
\end{equation}

Note that Eq.~\eqref{quadratic} describes the case of the Kitaev chain with, in general, inhomogeneous couplings but with the fermion pairing and hopping terms being of equal strength \cite{K01}. The model has time reversal, particle-hole, and chiral symmetry and thus belongs into the BDI symmetry
class in the classification of symmetry protected topological order \cite{Schnyder_2008}. In its topological phase, the model has two Majorana zero modes. This Hamiltonian with inhomogeneous couplings can now be diagonalized, for example, by building
raising and lowering operators as a
linear combination of the Majorana modes $\psi_\ell$, that is,
\be\label{linear}
&&\Psi_k = \sum_{\ell=1}^{L} \alpha_{k,\ell}\psi_{2\ell-1}+i\beta_{k,\ell}\psi_{2\ell}\,,\non\\
\quad
&&\Psi_k^\dagger = \sum_{\ell=1}^{L} \alpha_{k,\ell}\psi_{2\ell-1}-i\beta_{k,\ell}\psi_{2\ell}\,,
\ee
where $k=1,\dots,L$ and the
wave functions $\alpha_{k,\ell}$
and $\beta_{k,\ell}$ are recalled in App.~\ref{sec:standardd}. 
The raising and lowering operators $\Psi_k^\dagger$ and $\Psi_k$
form again a complex fermionic
algebra
and the Hamiltonian becomes diagonal in these eigenmodes
\be
H = \sum_{k=1}^{L} \epsilon_k \left[\Psi_k^\dagger,\Psi_k\right]\,.
\ee
The quasienergies $\epsilon_k$ are the roots of the characteristic
polynomial of a certain antisymmetric tridiagonal matrix,
as emphasized in \cite{K01}.
All the $2^L$ eigenvalues of (\ref{quadratic}) are given by,
\be\label{Efree}
E=\pm \epsilon_1 \pm \epsilon_2 \pm \cdots \pm \epsilon_L\,.
\ee
This spectral decomposition characterizes free-fermionic models. We recall that the linear transformation (\ref{linear})
between the physical modes $\psi_\ell$ and the eigenmodes
$\Psi_k$ and $\Psi_k^\dagger$ is essential to compute physical quantities. In fact,
Wick's theorem can be used to express expectation values of strings of Majoranas
in terms of two-point correlators \cite{SML64}.

Recently,
an intriguing free fermionic model
which does not have the form (\ref{quadratic}) was introduced
and solved \cite{F19}. Fendley's Hamiltonian contains three-spin interactions and is given by
\be\label{HF}
H_F=-\sum_{\ell=1}^{L-2}\lambda_\ell \sigma_{\ell}^x\sigma_{\ell+1}^z\sigma_{\ell+2}^z\,.
\ee
Applying the Jordan-Wigner transformation \eqref{JW}, one obtains the following 4-fermion Hamiltonian,
\begin{eqnarray}
\label{HF4}
H_F&=&-\sum_{\ell=1}^{L-2}\lambda_\ell(1-2c_\ell^\dagger c_\ell)(c^\dagger_{\ell+1}-c_{\ell+1})(c^\dagger_{\ell+2}+c_{\ell+2}) \nonumber \\
&=&\sum_{\ell=1}^{L-2}\lambda_\ell\psi_{2\ell-1}\psi_{2\ell}\psi_{2\ell+2}\psi_{2\ell+3}\,.
\end{eqnarray}
We see that, when written 
in terms of complex fermions, we have again a nearest-neighbor hopping and pairing between sites $\ell+1$ and $\ell+2$ as in the transverse Ising chain but now the sign of these terms does depend on the occupation of site $\ell$. Surprisingly, despite being composed of 4-fermion terms, the spectrum of (\ref{HF}) has the same free fermionic form (\ref{Efree}),
but with quasienergies given by the roots of a polynomial generated by a third order
recurrence relation \cite{F19}. We should remark here that the Fendley model is different from the three-spin extension of the transverse-field Ising model which has been studied already earlier in a different context \cite{Kopp_2005,Niu_2012}. In the latter case, the three-spin term has the form $\sigma^z_\ell \sigma^x_{\ell+1}\sigma^z_{\ell+2}=(c_\ell-c_\ell^\dagger)(c_{\ell+2}+c_{\ell+2}^\dagger)=\text{i}\psi_{2\ell}\psi_{2\ell+3}$, i.e., it remains bilinear in the fermionic operators. The beautiful
solution of the model \eqref{HF} in Ref.~\cite{F19} exploits
the fact that the Hamiltonian is the sum of local energy density operators that are generators of an elementary
algebra. This allows one to use integrability to express (\ref{HF}) in the diagonal form
\be
H_F =\sum_{k=1}^{\lfloor L/3\rfloor} \epsilon_k \left[\mathcal{F}_k^\dagger,\mathcal{F}_k\right]\,,
\ee
where the operators $\mathcal{F}_k$ and $\mathcal{F}_k^\dagger$
form a complex fermionic algebra as in
(\ref{fermions}) with $\lfloor x \rfloor$
denoting the floor function of $x$.
The operators $\mathcal{F}_k$ and $\mathcal{F}_k^\dagger$ are given in terms of an appropriate transfer matrix
and an edge operator \cite{F19}.
We stress here that the Hamiltonian (\ref{HF}) has open boundary conditions; solving the periodic chain remains an open problem. An important aspect of Fendley's model is that the physical
modes cannot be linearly expressed in terms of $\mathcal{F}_k^\dagger$ and $\mathcal{F}_k$, in contrast to (\ref{linear}).
As a consequence,
Wick's theorem cannot be applied, and correlations are difficult to compute.
For a characterization of free fermionic models solvable by generator to generator
maps, see also Ref.~\cite{CF20}.

The work \cite{F19} has motivated a number of further developments.
In references \cite{AP20a,AP20b} a family of models generalizing (\ref{HF})
to multispin interactions and also to free Z(N) parafermionic degrees
of freedom \cite{B89,F13} was introduced and its critical
behavior was analyzed.
It was further shown that these models, in the fermionic case,
belong to a class of
Hamiltonians with a certain frustration graph \cite{ECF20}. A powerful method
to analyze the spectral gap, including the cases where the model 
includes quenched disorder,
was proposed in \cite{AHP21}. For the parafermionic case, a spectral correspondence
with XY quantum chains with multispin interactions was argued in \cite{AP21}.
In Ref.~\cite{YM22},
considering multispin chains with periodic
boundary conditions, connections with the generalized Onsager algebra \cite{S20}
and generalized Yang-Baxter algebra \cite{GP21} were established.

Interestingly, all these multispin models with free fermionic spectra have a multicritical point
where the gap vanishes with dynamical critical exponent $z=(p+1)/2$ for
positive integer $p$ while the energy density operator
of these models typically acts on $p+1$ lattice sites. The Hamiltonian (\ref{HF}), for example, has $p=2$ and therefore $z=3/2$. We recall that many known critical spin chains have conformal symmetry ($z=1$) although ferromagnetic models with $z=2$ and spin chains such as the Fredkin and Motzkin models with multiple dynamics corresponding to different $z$ values are also known \cite{Chen_2017}. We also note that it has been recently suggested that the spin-1/2 Heisenberg chain shows superdiffusion with $z=3/2$ \cite{2022arXiv221117181N}. However, these models are typically very difficult to investigate analytically. Therefore, the multispin free fermionic
models might allow to consider dynamics in critical chains beyond the CFT regime in more detail and rigour. Unfortunately, despite the free fermion structure of the spectrum, the impossibility of using Wick's theorem makes the calculation of
correlation functions a difficult task even in this case. In addition, we should mention that numerical results are also difficult to obtain because the latter models typically have a spectrum with a global degeneracy that grows exponentially with the lattice size
(see Sec.~\ref{sec:alg} for details).

In this context, a question that can be asked is whether or not there are `standard' free fermionic
spin chains
with the same spectrum as the Fendley model \cite{F19} or its extensions \cite{AP20a,AP20b},
except for the exponential degeneracy. If such models can be constructed, then one can use Wick's theorem to analyze correlations and to perhaps grasp some universal physical behavior of the multispin chains beyond the spectral level.

This question 
can also be
put in the following way: is there a tridiagonal antisymmetric matrix whose
characteristic polynomial yields the quasienergies $\epsilon_k$ of
the generalized free fermionic quantum
chains? The answer to this question is affirmative and, as a matter of fact,
this problem has been previously considered in the literature
\cite{F90,S93}. In reference \cite{S93}, an algorithm which provides
the tridiagonal matrix associated with a given polynomial is given.
Therefore, when applied to the characteristic polynomial whose roots are the quasienergies $\epsilon_k$,
we actually obtain a set of couplings forming a standard free fermionic chain with the same spectrum
as the multispin models. 
In general though, the obtained couplings are inhomogeneous. 

The idea of reconstructing a tridiagonal matrix from the spectral data has been used
to construct quantum XY spin chains with perfect state transfer, see \cite{VZ12} and references therein.
In this case, the obtained spin chains are also inhomogeneous. Here, instead
of imposing perfect state transfer, we require the spectrum to be that of the 
free-fermionic multispin models.
Furthermore, the entanglement entropy of inhomogeneous XY chains was recently studied in
the framework of orthogonal polynomials \cite{CNV19,FGL21} while higher-dimensional cases were
considered in \cite{BCNPAV22}. Yet another interesting example of inhomogeneous chains is the so called rainbow chain \cite{BSRS} which is also of the XY type. In addition of shedding light on the multispin chains, the models we construct here also contribute to the study of inhomogeneous models in general. Let us finally 
mention that the eigenenergies  of XY (or XX) models are formed 
by the composition of two decoupled Ising chains. We can therefore produce 
inhomogeneous XY toy models using the same algorithm.

This paper is organized as follows: In Sec.~\ref{sec:multi}, we recall some basic facts
about the multispin free fermions, including the underlying exchange algebra and the polynomials
which fix the quasienergies of the system. The quasienergies are studied numerically and zero modes
are found within certain regimes of the coupling parameters. In Sec.~\ref{sec:inhomo}, inhomogeneous
quantum Ising chains are constructed based on the Schmeisser algorithm. In Sec.~\ref{sec:corr},
we compute correlations for the inhomogeneous models and uncover its phase diagram,
which is found to be dependent on certain integers related to the number of energy density operators defining
the Hamiltonian. We discuss the obtained results and perspectives in Sec.~\ref{sec:conclu}.
In App.~\ref{sec:standardd}, to make the paper self contained, we briefly review some aspects
of the solution of inhomogeneous quantum Ising chains.

\section{Multi-spin free-fermionic models}\label{sec:multi}
In this section, we present
some important properties
of the multispin free fermionic models \cite{F19,AP20a,AP20b}.
\subsection{Algebra and polynomial}\label{sec:alg}
The Hamiltonian is given by a sum of $M$ energy 
density operators
$h_\ell$,
\be\label{ham}
-H=\lambda_1h_1 + \lambda_2h_2 + \cdots +\lambda_Mh_M\,,
\ee
that are the generators of
the following algebra,
\be\label{alg}
h_\ell h_{\ell+1} &=& - h_{\ell+1}h_\ell\,,\non\\
h_\ell h_{\ell+2} &=& - h_{\ell+2}h_\ell\,,\non\\
 &\vdots& \non\\
h_\ell h_{\ell+p} &=& - h_{\ell+p}h_\ell\,,\non\\
\left[h_\ell,h_{\ell'}\right]&=& 0 \quad \text{if} \quad |\ell-\ell'|>p\,,
\non\\
h_{\ell}^2&=&1\,,
\ee
with $p$ a positive integer. We consider nonnegative real couplings $\lambda_\ell$. Using the method of Ref.~\cite{F19}, one can build
a set of conserved charges from products of the generators 
$h_\ell$, and use integrability to derive the spectrum of (\ref{ham}) independent of the
representation of the algebra, up to possible zero modes. The Hamiltonian (\ref{ham}) has
a free fermionic spectrum given by
\be\label{spec}
E=\pm \epsilon_1 \pm \epsilon_2 \pm \cdots \pm \epsilon_{\bar M}
\ee
where the quasi-energies $\epsilon_j$ are related to the the roots $z_j$ of the polynomial,
\be\label{recpol}
P_M(z) = P_{M-1}(z) - z\lambda_M^2P_{M-(p+1)}(z)
\ee
by
\be\label{eps}
\epsilon_j = \frac{1}{\sqrt{z_j}}\quad \text{with} \quad
\overline{M} = \left\lfloor \frac{M+p}{p+1} \right\rfloor\,.
\ee
The initial conditions are $P_\ell(z)=1$ if $\ell\leq 0$. Also, we assume $\epsilon_1<\epsilon_2<\cdots <\epsilon_{\bar M}$.
Explicitly, the polynomial is given by
\be\label{pol}
P_M(z) = \sum_{l=0}^{\overline{M}} C_M^{(l)}(-z)^l
\ee
with coefficients
\be\label{polcoef}
C_M^{(l)} = \sum_{j_1=1}^M\sum_{j_2=j_1+p+1}^M
\!\!\cdots\!\! \sum_{j_{l}=j_{{l}-1}+p+1}^M \lambda_{j_1}^2\lambda_{j_2}^2\dots \lambda_{j_l}^2 \,.
\ee
In passing, we want to mention that, remarkably, the polynomials (\ref{recpol}) in the homogeneous
case $\lambda_\ell=1$ have appeared in classical Rydberg blockade models \cite{LY22} and in
the enumeration of open walks of fixed length and algebraic area on a square lattice \cite{OP21}. For the case $p=1$, the algebra (\ref{alg}) was used to construct
an algebraic generalization of the Jordan-Wigner transformation \cite{M16}
and to make various connections to the Onsager algebra \cite{M21}.

One can consider different representations of the algebra (\ref{alg}), see Ref.~\cite{AP20b} for details.
For example,
\be\label{hp}
h_\ell =\sigma_{\ell}^x\sigma_{\ell+1}^z\cdots\sigma_{\ell+p-1}^z\sigma_{\ell+p}^z\,,
\ee
is a representation in $\mathbb{C}^{\otimes M+p}$. For this representation,
all energy states have the same exponential
degeneracy $2^{\lfloor p(M+p+1)/(p+1)\rfloor}$. To our knowledge, representations of the algebra (\ref{alg}) in $\mathbb{C}^{\otimes \bar M}$ for $p>1$ are not known.
The energy density
(\ref{hp}) usually contains multiple Majoranas. For example, we have
\begin{eqnarray}
 \label{examples}
&& h_\ell=\psi_{2\ell-1}\psi_{2\ell}\psi_{2\ell+2}\psi_{2\ell+3} \qquad\text{for } p=2, \\
&& h_\ell=-i\psi_{2\ell-1}\psi_{2\ell}\psi_{2\ell+2}\psi_{2\ell+3}\psi_{2\ell+6}\psi_{2\ell+7} \qquad\text{for } p=4 \nonumber \, .
\end{eqnarray}
For odd values of $p$, the representation (\ref{hp}) with the Jordan-Wigner transformation (\ref{JW}) leads to energy densities with an ever increasing number of Majoranas. This can be cured by considering rotated versions of Eq.~(\ref{JW}).

In the case $p=1$, the algebra (\ref{alg}) admits an Ising representation
\be\label{isingrep}
&&h_{2\ell-1}=\sigma_\ell^x\quad \text{for}\quad\ell=1,\dots,L\,,
\non\\
&&h_{2\ell}=\sigma_\ell^z\sigma_{\ell+1}^z\quad \text{for}\quad\ell=1,\dots,L-1\,,
\ee
with $M=2L-1$ generators. We can equivalently write $h_{2\ell-1}=i\psi_{2\ell-1}\psi_{2\ell}$
and 
$h_{2\ell}=i\psi_{2\ell}\psi_{2\ell+1}$ using Eq.~(\ref{JW}).
As recalled in App.~\ref{sec:standardd}, one can associate a tridiagonal matrix with this
representation, see Eq.~(\ref{Hpvec}) with $w_j\rightarrow \lambda_j$, to obtain the single-particle energies $\varepsilon_k$ which yield the many-body spectrum given by 
Eq.~\eqref{Efree}. 
The even $M$ case is slightly more subtle. We can eliminate one generator, say $h_1=\sigma_1^x$,
producing an even number $M=2L-2$ of terms. In the spin representation, Eq.~\eqref{isingrep}, the Hamiltonian then commutes with $\sigma_1^z$ leading to a block structure of the Hamiltonian and a double degeneracy of the entire many-body spectrum. In the Majorana language, this is equivalent to removing one line and one column of the tridiagonal matrix in  (\ref{Hpvec}). As a consequence, the tridiagonal matrix acquires an odd dimension and therefore a null eigenvalue $\varepsilon_1=0$. This zero eigenvalue in the single-particle spectrum is then responsible for the double degeneracy in the many-body spectrum \eqref{Efree}. One could instead also redefine the generator $h_2=\sigma_1^z\sigma_2^z\rightarrow h_2=\sigma_2^z$ and the algebra (\ref{alg}) would still be satisfied. While this would kill the extra degeneracy, it would introduce a trilinear term $\sigma_2^z=-i\psi_1\psi_2\psi_3$ to the Hamiltonian and 
one would no longer be able to associate a tridiagonal matrix to it.

\begin{figure}
    \includegraphics[width=0.99\columnwidth]{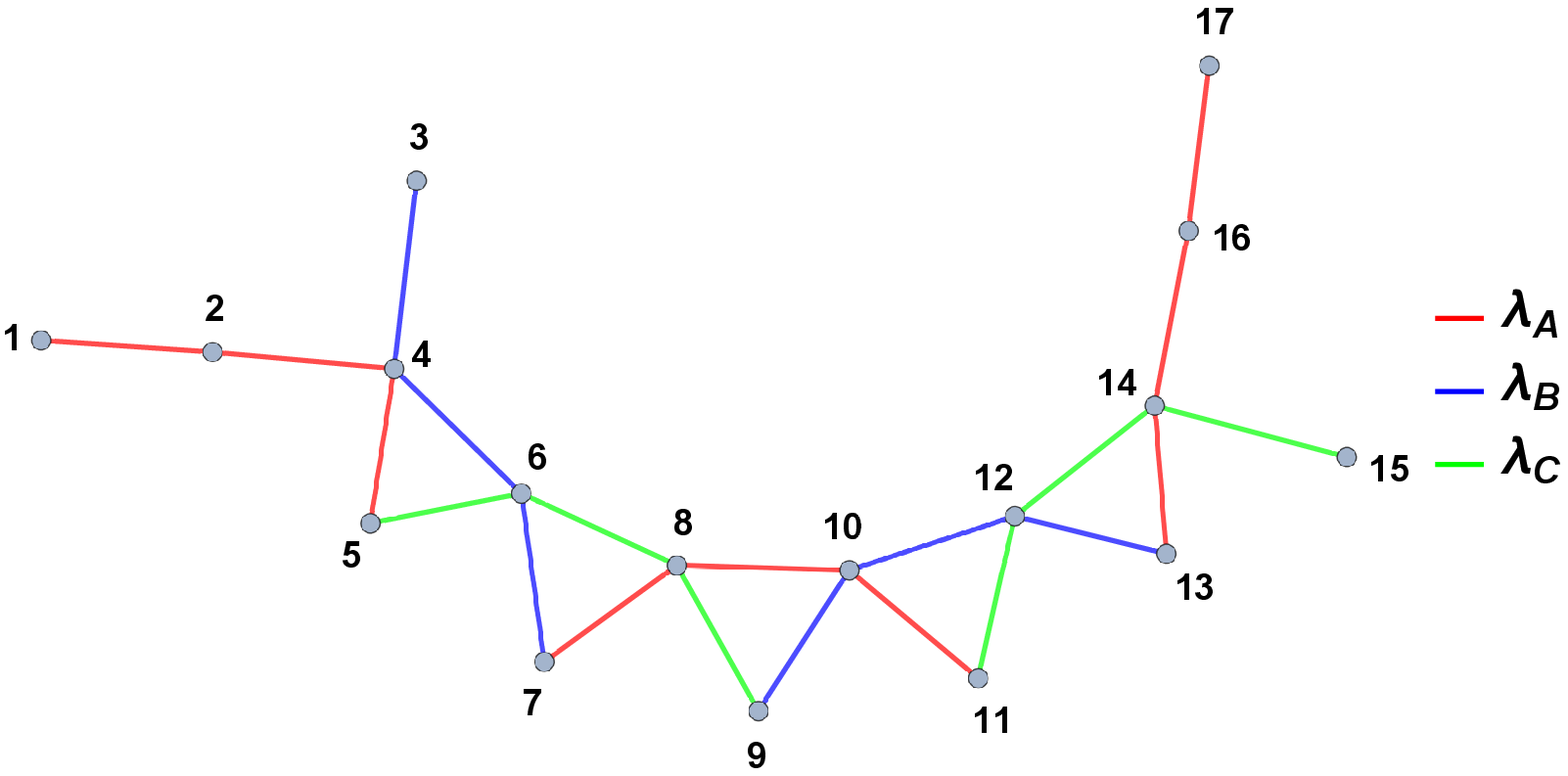}
    \includegraphics[width=0.99\columnwidth]{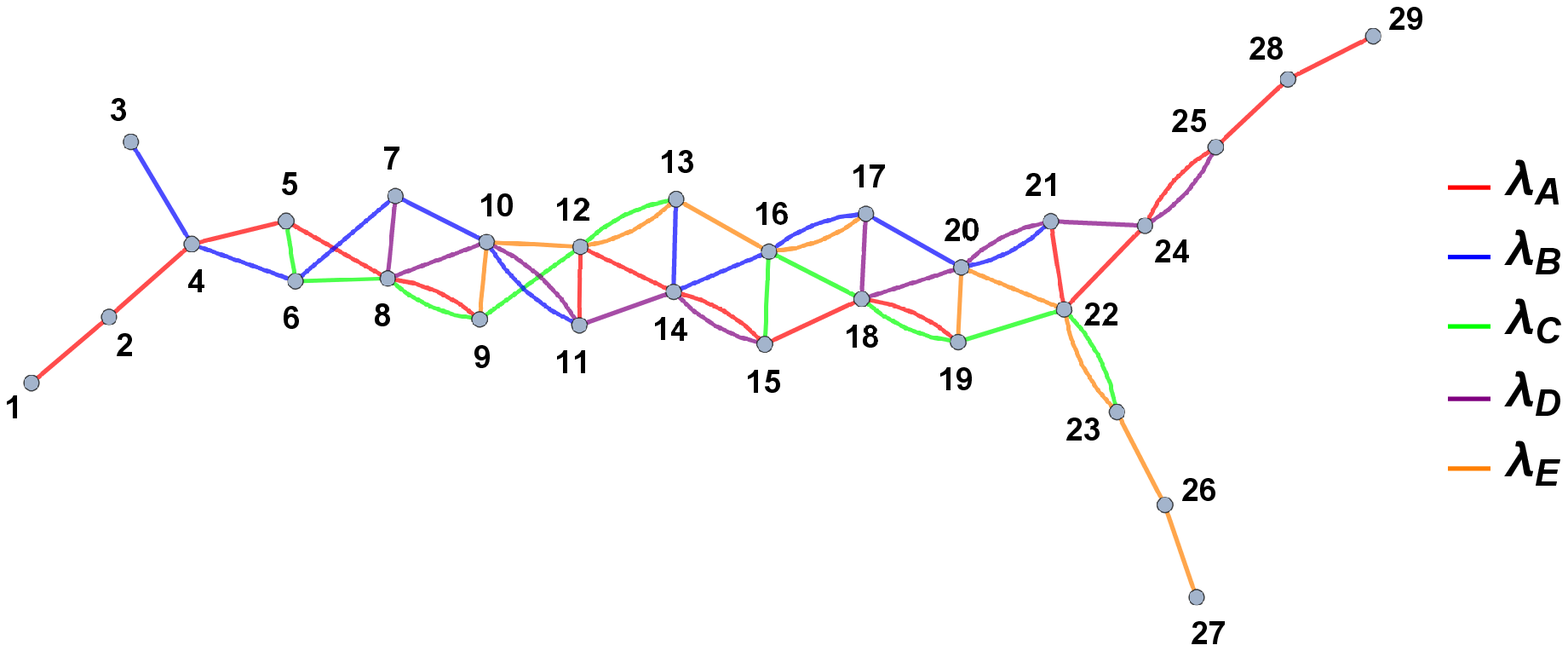}
\caption{Graphs illustrating the Hamiltonian (\ref{ham}) with local terms (\ref{examples}) in the Majorana language. Each vertex contains a Majorana while the edges with different colors represent the couplings split according to their index parity. The case $p=2$ with $M=7$ is shown on the top and the case $p=4$ with $M=11$ on the bottom.}
    \label{fig:Majgraph}
\end{figure}

There is a natural splitting of the Hamiltonian (\ref{ham}) according to the
``parity''
\be\label{parity}
\mathfrak{p}_{\ell}=\ell \bmod (p+1)\,,
\ee
of the index $\ell$ of the parameters $\lambda_{\ell}$ in Eq.~\eqref{ham}. Although the spectrum of this Hamiltonian can be computed for arbitrary couplings $\lambda_{\ell}$, one
convenient way to analyze the phase diagram is to
consider the case where the couplings with the same parity $\mathfrak{p}_{\ell}$ are the same. For $p=1$,
this means that we can write (\ref{ham})
as
\be\label{Hp1}
-H=\lambda_A H_A+\lambda_B H_B\,,
\ee
with
\be
H_A=\sum_{\ell=1}^{\lfloor (M+1)/2\rfloor} h_{2\ell-1}\,,\quad
H_B = \sum_{\ell=1}^{\lfloor M/2\rfloor} h_{2\ell}\,, 
\ee
while for $p=2$ we have,
\be\label{Hp2}
-H=\lambda_A H_A+\lambda_B H_B+\lambda_C H_C\,,
\ee
with
\be
&&H_A=\sum_{\ell=1}^{\lfloor (M+2)/3\rfloor} h_{3\ell-2}\,,\quad H_B=\sum_{\ell=1}^{\lfloor (M+1)/3\rfloor} h_{3\ell-1}
\,,\non\\&& H_C = \sum_{\ell=1}^{\lfloor M/3\rfloor} h_{3\ell}\,,
\ee
and similarly for higher values of $p$. This splitting was first proposed in Ref.~\cite{F19} to analyze the phase diagram. Recently,
it was shown, furthermore, that in the case of periodic boundary conditions the operators $H_{A,B,...}$ form a generalized
Onsager algebra \cite{YM22} if $\mathfrak{p}_M=0$. As an illustration, we show in Fig.~\ref{fig:Majgraph} graphs
of the Hamiltonians \eqref{examples} in the Majorana language with split couplings for $p=2$ and $p=4$.

\subsection{Quasienergies and zero modes}\label{sec:zeromodes}
In this section, we numerically investigate the roots of the polynomial
(\ref{pol}) in the scenario of split couplings. The roots of the polynomial (\ref{pol}) in general do not have a closed formula for $p>1$,
except at the multicritical point $\lambda_{j}=1$ in the bulk limit when $\mathfrak{p}_M=0$
\cite{F19,AP20a,AP20b}.

The largest root $z_\text{max}$ which gives the
smallest quasienergy $\epsilon_1=1/\sqrt{z_\text{max}}$ is particularly important. Two
scenarios do occur. Firstly, $\epsilon_1$ may have a finite value in the thermodynamic limit. Therefore a gap, $\Delta=2\epsilon_1\neq 0$, between the many-body ground state state and the first excited state exists. Alternatively, $\epsilon_1$ might be exponentially small with the system size, implying that the many-body ground state is degenerate in the thermodynamic limit. These exponentially small eigenvalues and the associated degeneracy of the ground state in the thermodynamic limit are of topological nature. This emerging degeneracy should not be confused with  the global  built-in degeneracy 
that grows with the lattice size and that is
independent of the couplings of
the Hamiltonian.
The global degeneracy arises 
as a consequence of single particle eigenvalues which are exactly zero and associated with a given representation of the algebra (\ref{alg}). These trivial, exactly zero, eigenvalues are not considered in the following.

Interestingly, the behavior of the lowest quasienergy $\epsilon_1$ depends on the 
parity 
(\ref{parity}) of the number of generators $\mathfrak{p}_M$.

Let us first recall the simplest case $p=1$, for which the quasienergies $\epsilon_k$ are analytically known
in terms of the transverse field $\lambda\equiv \lambda_A/\lambda_B$. They are given by
\be\label{ek}
\epsilon_k = \sqrt{1+\lambda^2+2\lambda \cos\mathfrak{q}_k}\,,\quad k=1,\dots,\bar M,
\ee
where for even $M$ ($\mathfrak{p}_M=0$) we have \cite{AP20b},
\be
\mathfrak{q}_k=\frac{2\pi k}{M+2}\,,
\ee
while for odd $M$ ($\mathfrak{p}_M=1$), $\mathfrak{q}_k$ is a solution of the transcendental
equation \cite{P70}
\be
\lambda\sin\left(\frac{M+3}{2}\mathfrak{q}_k\right)=-\sin\left(\frac{M+1}{2}\mathfrak{q}_k\right)
\,.
\ee
As an aside, we note here that quite similarly the quasienergies of
the SSH chain \cite{SMKS14} can also be obtained without
resorting to a transcendental equation when the number of lattice sites is odd.

\begin{figure}[h!]
\centering
\includegraphics[width=0.99\columnwidth]{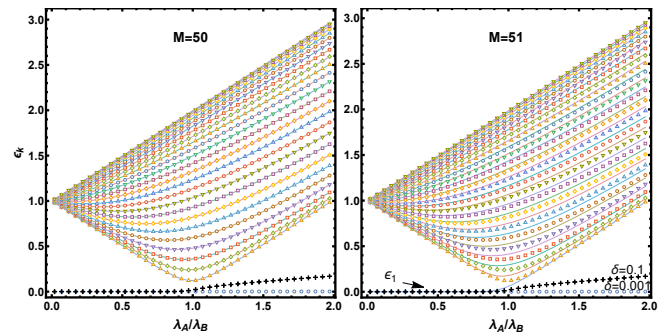}
    \caption{Positive part of the symmetric spectrum $\epsilon_k$ as a function of the transverse field $\lambda=\lambda_A/\lambda_B$ for $M=50$ and $M=51$ using Eq.~\eqref{ek} (solid lines). The points are the same in both pictures and correspond to a numerical solution of the case $M=51$ with a defect $\delta=10^{-3}$. For the quasienergy $\epsilon_1$ only, also an intermediate defect strength $\delta=10^{-1}$ is shown (black crosses).}
    \label{fig:pic2}
\end{figure}

In Fig.~\ref{fig:pic2}, we plot the positive quasienergies (\ref{ek}) for $M=50$ and $M=51$ obtained from the roots of the polynomial \eqref{pol}. For odd $M$,
we can observe the expected emergence of a zero mode for $\lambda < 1$ 
(marked as $\epsilon_1$ in the right panel of Fig.~\ref{fig:pic2}). 
On the other hand, for even $M$, there is no non-trivial zero mode arising as a consequence of the transverse field $\lambda$.

We remark, however, that exact zero modes can always be added to (\ref{ek}). We can, for example, consider the polynomial for  $M$ even as being the limit of the odd $M+1$  
case introducing a surface defect in the quantum Ising representation (\ref{isingrep}).
More specifically, we set
the coupling parameter for the first site
of the chain as $\lambda\sigma_1^x\rightarrow \lambda\delta\sigma_1^x$ where
$\delta$ is the defect parameter. Then, by varying $\delta$, we move from the odd $M+1$ case ($\delta=1$) to the even $M$ case ($\delta=0$). As shown in Fig.~\ref{fig:pic2}, the defect in the limit $\delta\to 0$ indeed leads to a trivial zero mode independent of $\lambda$. In the figure, we
plot the numerical result (points) together with the solutions of Eq.~\eqref{ek} (continuous lines). While we are only interested in the limits $\delta=0$
and $\delta=1$ here, we point out that a defect with arbitrary $\delta$ leads
to interesting boundary phenomena in the quantum Ising chain, see for example Ref.~\cite{FAGP16}.

We now move to the case $p=2$, for which we need to consider three different parities $\mathfrak{p}_M=0,1,2$ and analyze the behavior of the energies $\epsilon_k$
as a function of $\lambda_{A,B,C}$. In Ref.~\cite{F19}, the case $\mathfrak{p}_M=0$ was considered and it was argued that
the phase diagram as a function of $\lambda_A/\lambda_C$ and $\lambda_B/\lambda_C$
is divided into three gapped phases ($\Delta=2\epsilon_1 \neq 0$), separated by critical lines
($\Delta\sim 1/M^z$)
with dynamical exponent $z=1$ which meet at the multicritical point
$\lambda_A=\lambda_B=\lambda_C$ with
$z=3/2$.

\begin{figure}[]
\centering
\includegraphics[width=8.5cm]{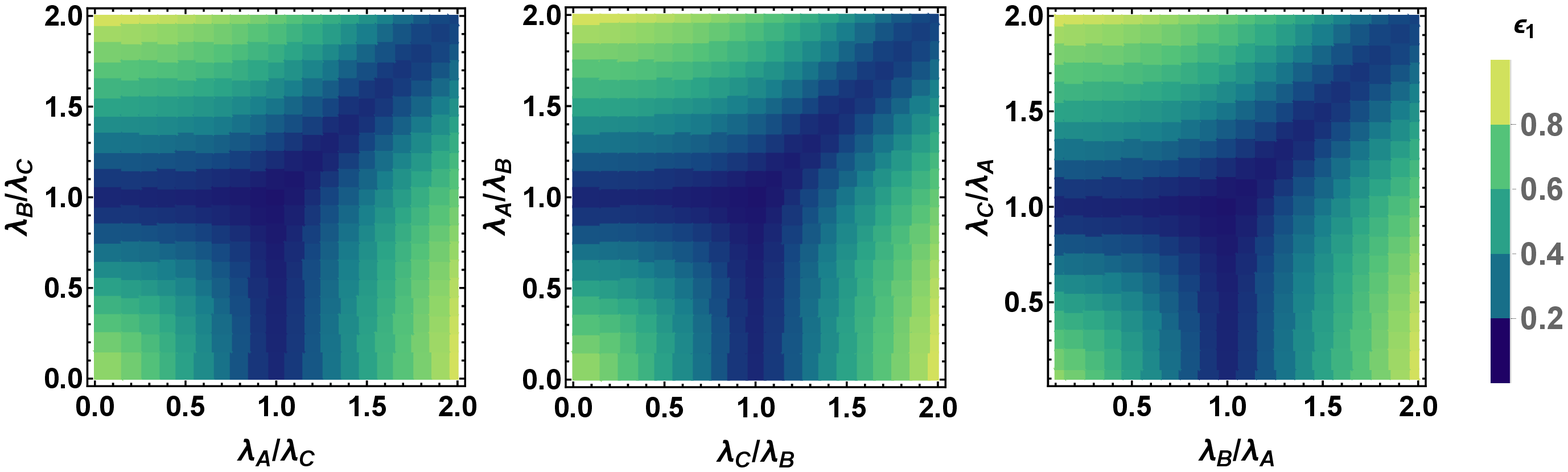}
(a) $M=99$
\includegraphics[width=8.5cm]{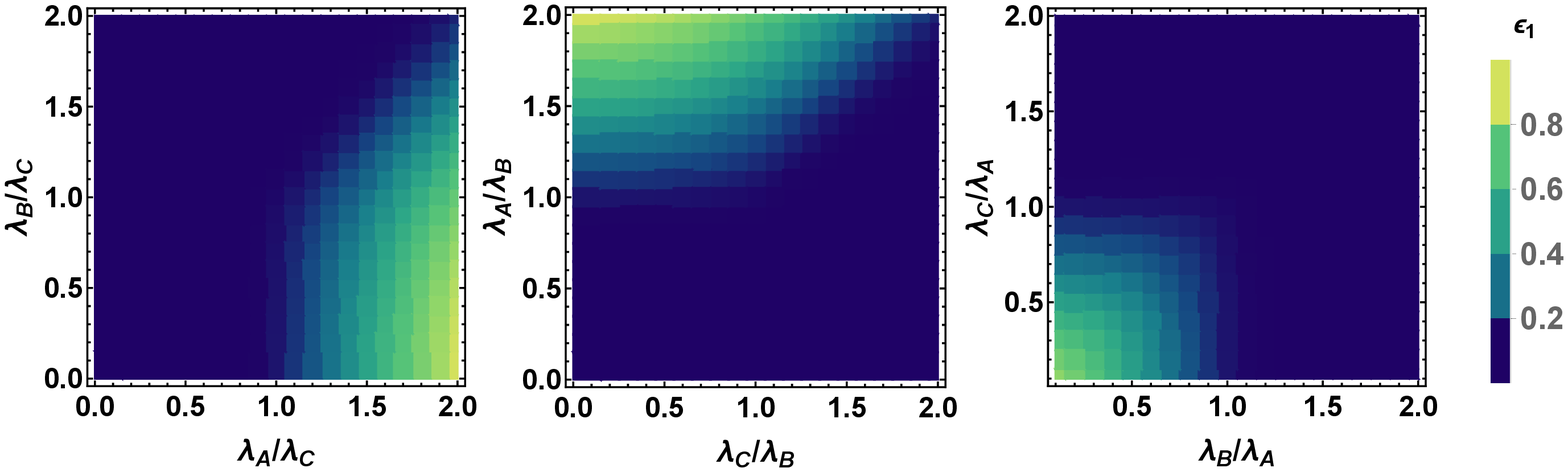}
(b) $M=100$
\includegraphics[width=8.5cm]{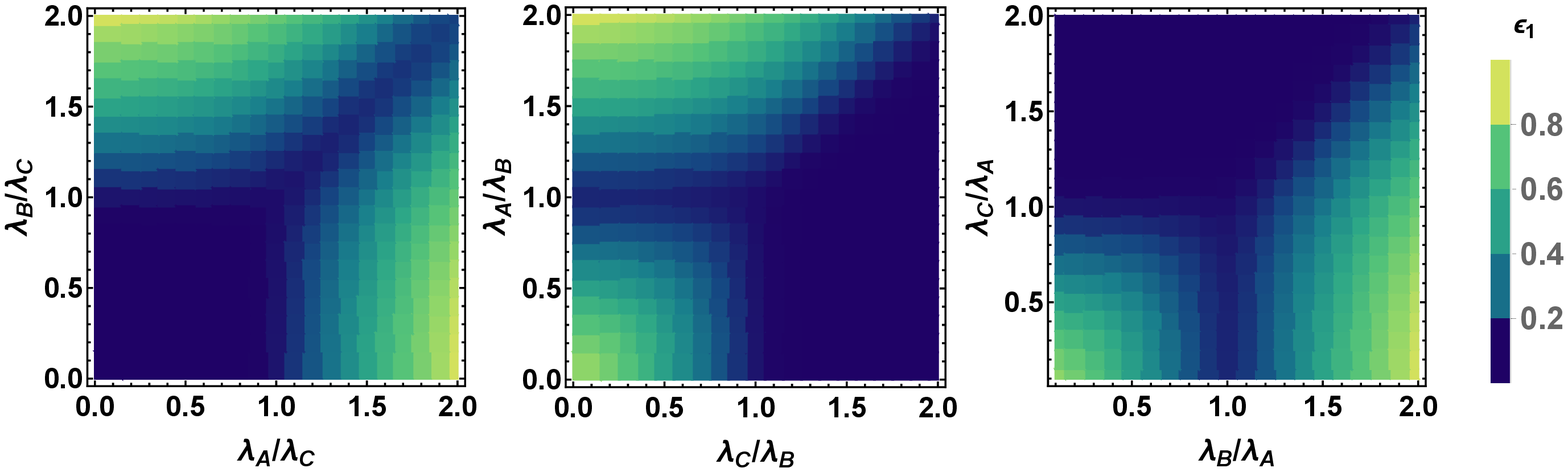}
(c) $M=101$
    \caption{Smallest positive quasienergy $\epsilon_1$ for $M=99,100,101$ (corresponding, respectively, to the parities $\mathfrak{p}_M=0,1,2$), as a function of ratios of the coupling parameters $(\lambda_A,\lambda_B,\lambda_C)$.
    The values in the dark regions are non-zero but extremely small and go to zero in the thermodynamic limit.}\label{fig:pic3}
\end{figure}

Considering all possible parities, we find numerically that two non-trivial zero modes can be present for the parities $\mathfrak{p}_M=1,2$, see Fig.~\ref{fig:pic3}. Here the smallest positive quasi-energy $\epsilon_1$ (the specrum is symmetric) is shown for $M=99$ ($\mathfrak{p}_M=0$),
$M=100$ ($\mathfrak{p}_M=1$) and $M=101$ ($\mathfrak{p}_M=2$) as a function
of the split coupling parameters. We used a grid $0.1\leq \lambda_j \leq 2$ for every split coupling $\lambda_j$ and a step size $\delta\lambda_j=0.1$. For $M=99$, no zero modes exist. In this case, previously considered in Ref.~\cite{F19},
the quasienergy $\epsilon_1\neq 0$ leads to a finite gap $\Delta=2\epsilon_1$ between the many-body ground state and the first excited state in the three regions
separated by the critical lines. For $M=100$, we observe two regions with zero modes, while
for $M=101$ only one region has zero modes. In these cases, 
similarly to the ordered phase of the $p=1$ case,
the gap in the thermodynamic limit between the degenerate ground state and first excited state is given
by the second smallest quasienergy $\epsilon_2$, that is, $\Delta=2\epsilon_2\neq 0$. 

\begin{figure}[htp!]
\centering
\includegraphics[width=8.5cm]{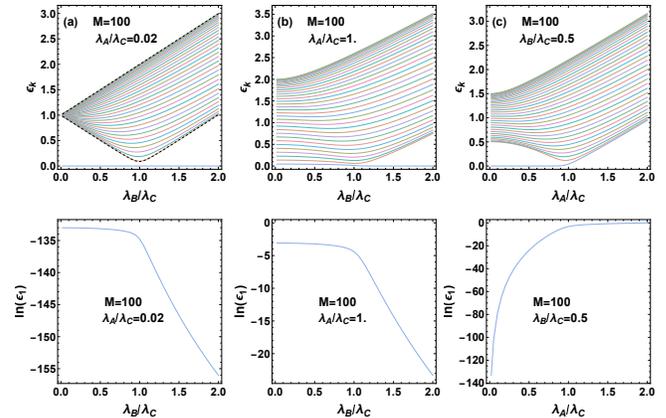}
\caption{Representative profiles of the quasienergies for $p=2$. The dashed lines in panel (a) correspond to Eq.~(\ref{Ansatz}). The bottom panels show the energy dependence of what will become the zero mode in the thermodynamic limit on the coupling parameters.
Note that for the configuration in panel (c) the zero mode is only present for $\lambda_A/\lambda_C<1$.}\label{fig:quasip2}
\end{figure}

We note that the 
bulk modes $\epsilon_{k\neq 1}$ are expected to be independent of the parity of the number of generators in the thermodynamic limit consistent with our numerical findings. 
Some representative cuts obtained by fixing one of the ratios of the coupling parameters are show in Fig.~\ref{fig:quasip2}. In panel (a), we fix a $\lambda_A/\lambda_C=0.02$ and compute
$\epsilon_{k}$ as a function of $\lambda_B/\lambda_C$. In this limit, we observe, as expected, a similar behavior as in the
$p=1$ case. The dashed black line is an ansatz based on (\ref{ek}), namely,
\begin{equation}
\label{Ansatz}
\epsilon_k = \sqrt{1+(\lambda_B/\lambda_C)^2+2(\lambda_B/\lambda_C) \cos\left(\frac{\pi k}{\bar{M}+1}\right)}
\end{equation}
for $k=1$ and $k=\bar M$, and we observe an excellent agreement. As we move away from 
$\lambda_A/\lambda_C \approx 0$, see panels (b) and (c), the profile of the quasienergies clearly changes.

We remark that the presence of zero
modes indicates that the system has non-trivial topological order. However, defining and computing any type of order parameter
for $p>1$ is a hard task, since it is not known how to compute correlations
directly for the Hamiltonian (\ref{hp}). In Ref.~\cite{F19}, it is mentioned that
the expectation values $\langle H_{A,B,C} \rangle$ can be used as indicators of order. It is, however, important to keep in mind that these expectation values are not proper order parameters since they are never zero in the thermodynamic limit except when the associated coupling is set to zero. This is similar to the Ising case where $\langle H_{A} \rangle = \langle \sum_{\ell=1}^L \sigma_\ell^x \rangle$
 and $\langle H_{B} \rangle = \langle \sum_{\ell=1}^{L-1} \sigma_\ell^z\sigma_{\ell+1}^z \rangle$ are also not proper order parameters.

We end this section with a few remarks: (i) For values of $p>2$, the phase diagram is characterized by gapped phases separated by critical hyperplanes meeting at
a multicritical point with $z=(p+1)/2$ \cite{AP20a,AP20b}, see also Ref.~\cite{ECF20}. We
expect that depending on the number of generators, regions with non-trivial zero modes will also be present for $p>2$. We leave such an analysis for future studies. (ii) We have checked that the zero modes are robust against quenched disorder confirming their topological character. (iii) Finally, we have verified that the so called Laguerre bound \cite{AHP21},
which can be efficiently computed, 
gives a good approximation for the
smallest quasi-energy $\epsilon_1$ 
in the zero mode regions.

\section{Inhomogeneous Ising chain}\label{sec:inhomo}
In this section, we argue that it is always possible to construct a quantum Ising chain (\ref{Ising}), with in general inhomogeneous couplings $w_{\ell}$, which has the same spectrum as the multispin chain (\ref{ham}) for arbitrary $p$ and couplings $\lambda_{\ell}$. This is achieved by identifying the characteristic polynomial associated with the quantum Ising chain with a rescaled version of the polynomials $P_M(z)$ (\ref{pol}) of the multispin chain.

While the two chains will then, by construction, have the same spectrum and therefore the same partition function, their eigenstates will in general be different. Therefore correlation functions for the two chains will also be different. We find, however, that the inhomogeneous Ising chains constructed in this way have a bulk region in which the couplings tend to become homogeneous. We might then expect that inside these bulk regions at least some of the physical properties are the same as in the multispin chain. We are, in particular, interested in investigating possible multicritical points in these inhomogeneous spin chains to see whether or not the dynamical critical exponents $z>1$ are realized.
If they are, then these models could be useful to better understand dynamical critical behavior when $z>1$ because the Ising chains are bilinear in the Majorana operators and Wick's theorem therefore applies.

\subsection{Characteristic polynomial}\label{subsec:FSalgo}
As we briefly review in App.~\ref{sec:standardd}, the solution of the Hamiltonian (\ref{quadratic}) can be reduced to
the block diagonalization of the antisymmetric tridiagonal matrix $T_m$ with an odd
number of couplings $w_\ell$ ($\ell=1,\dots,2L-1$). The case with an even number
of couplings can be obtained by removing one row and one column.

We define this hopping matrix $T_m$
with dimension $(m+1)\times (m+1)$ by its elements
\be\label{hopa}
(T_m)_{ij} = w_i\delta_{j,i+1}-w_{i-1}\delta_{i,j+1}
\ee
where $i,j=1,\dots,m+1$ and $w_i$ are the $m$ coupling
parameters. In particular,
the quasi-energies $\epsilon_k$ entering the energy expression (\ref{Efree})
are the roots of the characteristic polynomial
\be\label{defWm}
W_m(x)=\det_{m+1}(T_m-x)
\ee
that is, $W_m(i\epsilon_k)=0$. Thanks to the tridiagonal form of $T_m$, the characteristic polynomial satisfies the following
recurrence relation
\be\label{recpolx}
W_m(x) = -xW_{m-1}(x)+w_m^2W_{m-2}(x)\,,
\ee
with the initial conditions $W_0(x)=-x$, $W_{-1}(x)=1$. When
$m$ is even, $x=0$ is a root of $W_m(x)$ and the many-body spectrum has a twofold degeneracy.

Note that the recurrence relation (\ref{recpolx})
is similar to the recurrence (\ref{recpol}) for $p=1$.
This motivates us to introduce the polynomials
\be\label{qpol}
Q_M(x) = x^{d(M)}P_M(-1/x^2)\,,
\ee
where
\be
d(M) = 2\overline{M} = 2\Big\lfloor{\frac{M+p}{p+1}}\Big\rfloor
\ee
is the degree of the polynomial $Q_M(x)$.
The polynomials $Q_M(x)$ have roots $i\epsilon_j =i/\sqrt{z_j}$
which are purely imaginary, since the roots $z_j$ are positive real numbers \footnote{This claim has been proved in the homogeneous case \cite{LY22} and checked numerically in the inhomogeneous case.}.
With this definition,
the polynomial $Q_M(x)$ has the same form as $W_m(x)$ up to the zero mode $x=0$ when $M$ is even.
Therefore, this definition does not include the built-in zero mode associated with
the even $M$ case, as discussed in Sec.~\ref{sec:zeromodes}. For other values of $p$,
we found that a possible definition of the Q-polynomial which would include zero modes when
$\mathfrak{p}_M\neq 1$ is
\be\label{qpolt}
\widetilde{Q}_M(x) = (-x)^{\widetilde{d}(M)}P_M(-1/x^2)\,,
\ee
where
\be
\widetilde{d}(M) =  2\Big\lfloor{\frac{M+p}{p+1}}\Big\rfloor + \Big\lceil \frac{M-1}{p+1}\Big\rceil-\Big\lfloor \frac{M-1}{p+1}\Big\rfloor\,,
\ee
with $\lceil x \rceil$ denoting the ceilling function of $x$. We have
\be
\widetilde{d}(M) - d(M)
&=&  \Big\lceil \frac{M-1}{p+1}\Big\rceil-\Big\lfloor \frac{M-1}{p+1}\Big\rfloor
\nonumber\\
&=&\begin{cases}
      0, & \text{if}\ \mathfrak{p}_M= 1 \\
      1, & \text{if}\ \mathfrak{p}_M\neq 1
    \end{cases}\,.
\ee

For concreteness, we consider in the following mainly the polynomial $Q_M(x)$ (\ref{qpol}),
although $\widetilde{Q}_M(x) $ (\ref{qpolt}) may appear as a limiting case of (\ref{qpol}).

The relation (\ref{recpol}) implies the following recurrence relation
for the Q-polynomials
\be
Q_M(x)&=&x^{2-2(\lceil \frac{M-1}{p+1}\rceil-\lfloor \frac{M-1}{p+1}\rfloor)}Q_{M-1}(x)
\non\\&+&\lambda_M^2Q_{M-(p+1)}(x)\,,
\ee
with the initial conditions $Q_j(x)=1$ if $j\leq 0$.

The central idea of this paper is that with a fine tuned
set of couplings $\{w_1,\dots,w_m\}$ we can take the polynomial $Q_M(x)$ (\ref{qpol})
as the characteristic polynomial of the hopping matrix $T_m$. Namely, we impose
\begin{equation}
\label{method}
W_{m}(x;\{w_1,\dots,w_{m}\}) = Q_M(x;\{\lambda_1,\dots,\lambda_M\}),
\end{equation}
which implies a set of equations relating the $w$-couplings with the $\lambda$-couplings.
Fixing the degrees of the polynomials implies
\be
m=d(M)-1=2\overline{M}-1\,,
\ee
which means that the number of $w$-couplings is less than $M$---the number of $\lambda$-couplings---except
for $p=1$ and odd $M$ when $d(M)=M+1$ and thus $m=M$. Solutions to (\ref{method}) are not
unique as can be easily checked for small values of $M$. For $p=1$, there is the obvious solution
$w_j=\lambda_j$ for all couplings.

Interestingly,
we found in the literature an algorithm that constructs a symmetric tridiagonal matrix out of
a given characteristic polynomial \cite{S93} as a solution to a problem previously proposed in
Ref.~\cite{F90}. We can easily adapt it to construct an
antisymmetric tridiagonal matrix. As a result, the algorithm iteratively finds the
$w$-couplings satisfying equation (\ref{method}). It can therefore be used to construct an inhomogeneous
quantum Ising chain which has the same spectrum as a multispin chain with a given $p$. We briefly describe this algorithm next.

\subsection{Schmeisser algorithm}\label{subsec:Salgo}
We reproduce here the algorithm by Schmeisser \cite{S93} for an
arbitrary monic polynomial $u(x)$ of degree $n$. For a
polynomial $g(x)$ of degree $k$ denote the coefficient of $x^k$ by $c(g)=a_k$.
\begin{algo*} (Modified Euclidean Algorithm \cite{S93}) For
\be
u(x)=x^n+a_{n-1}x^{n-1}+\cdots+a_0, a_\nu\in \mathbb{R}\non
\ee
with $\nu=0,1,\dots,n-1$, define
\be
f_1(x):=u(x),\quad f_2(x)=\frac{1}{n}u'(x)
\ee
and proceed recurrently as follows: If $f_{\nu+1}(x)\neq 1$, then by dividing $f_{\nu}$ by
$f_{\nu+1}$ with remainder $-r_{\nu}$, we obtain
\be
f_{\nu}(x)=q_{\nu}(x)f_{\nu+1}(x)-r_\nu(x) \, .
\ee
Now we define
\begin{enumerate}
\item[(i)] $c_\nu:=c(r_{\nu}),\quad f_{\nu+2}(x):=\frac{r_\nu(x)}{c_\nu},\quad\text{if}\quad r_\nu(x)\not\equiv0$\,,
\item[(ii)] $c_\nu:=0,\quad f_{\nu+2}(x):=\frac{f_{\nu+1}'(x)}{c(f_{\nu+1}')},\quad\text{if}\quad r_\nu(x)\equiv0$\,.
\end{enumerate}
If $f_{\nu+1}(x)\equiv 1$, we terminate the algorithm, defining $q_\nu(x):=f_\nu(x)$.
\end{algo*}

According to Ref.~\cite{S93}, the algorithm establishes the following connection
\be
u(x) = (-1)^n \det_{n}(A-x)
\ee
where $A$ is an $n\times n$ tridiagonal matrix with
elements
\be
A_{ij}=\sqrt{c_i}\delta_{j,i+1}+\sqrt{c_{i-1}}\delta_{i,j+1}-q_i(0)\delta_{i,j}\,.
\ee
We observe that the matrix $A$ is symmetric. In the cases where $q_i(0)=0$---which is the case for the Q-polynomials because they are polynomials in $x^2$---we can
construct an analog antisymmetric matrix by
\be
A_{ij}^{(as)}=\sqrt{c_i}\delta_{j,i+1}-\sqrt{c_{i-1}}\delta_{i,j+1}\,,
\ee
which has eigenvalues $i\epsilon_k$ with  real $\epsilon_k$.


\subsection{Application to the Q-polynomials}
We now apply the Schmeisser algorithm to the Q-polynomials (\ref{qpol}). As
described in Sec.~\ref{sec:multi}, these polynomials are characterized by the set of couplings
$\{\lambda_1,\dots,\lambda_M\}$ which are arbitrary in general. Here we will restrict ourselves again to the case where the couplings with the same parity are the same, see Eq.~(\ref{Hp1}) and Eq.~(\ref{Hp2}).

We are mostly interested in the behavior of these spin chains in the thermodynamic limit. The Schmeisser algorithm can be implemented efficiently, and we can run it for large values of $M$, $p$, and any couplings $\{\lambda_\ell\}$, followed by fixing $w_\ell=\sqrt{c_\ell}$. However, before doing so it is beneficial to first consider the simplest case of homogeneous couplings $\lambda_{A,B,C,...}=1$ and small values of $M$ for various $p$, see Table \ref{tab:1}. For example, the first line in Table \ref{tab:1}
means that
\be\label{Ising1}
H_{\rm Ising}^{\rm inhom}=-
\sqrt{\frac{3}{2}}\sigma_1^x-\sqrt{\frac{5}{6}}\sigma_1^z\sigma_2^z-\sqrt{\frac{2}{3}}\sigma_2^x
\ee
and the Hamiltonian (\ref{hp})
\begin{equation}
\label{xz1}
H_{XZ}^{(M=3)}(\{\lambda_\ell=1\})=
-\sigma_1^x\sigma_2^z-\sigma_2^x\sigma_3^z-\sigma_3^x\sigma_4^z\,,
\end{equation}
have the same spectrum up to degeneracies. For the $p=1$ considered here,
we have also the homogeneous Ising chain 
\be\label{Ising2}
H_{Ising}^{homogeneous}=-
\sigma_1^x-\sigma_1^z\sigma_2^z-\sigma_2^x
\ee
with the same spectrum as (\ref{Ising1}) and (\ref{xz1}). As another example, the $(p=2,M=6)$ line in Table \ref{tab:1}
means that
\be
H_{\rm Ising}^{\rm hom}=-\sqrt{3}\sigma_1^x
-\sigma_1^z\sigma_2^z
-\sqrt{2}\sigma_2^x\,,
\ee
and the Hamiltonian (\ref{hp}),
\begin{eqnarray}
\!\!\!\!\!\! H_{XZZ}^{(M=6)}&=&
-\sigma_1^x\sigma_2^z\sigma_3^z
-\sigma_2^x\sigma_3^z\sigma_4^z
-\sigma_3^x\sigma_4^z\sigma_5^z
\non\\
&&-\sigma_4^x\sigma_5^z\sigma_6^z
-\sigma_5^x\sigma_6^z\sigma_7^z
-\sigma_6^x\sigma_7^z\sigma_8^z,
\end{eqnarray}
have the same spectrum up to degeneracies.
\begin{table}[!]
\centering
\begin{tabular}{|c|c|c|c|}
\hline
 $p$ & $M$ & $Q_M(x)$ & \{$w_\ell$\} \\ \hline 
 1 & 3 & $1+3 x^2+x^4$ & $\left\{\sqrt{\frac{3}{2}},\sqrt{\frac{5}{6}},\sqrt{\frac{2}{3}}\right\}$ \\ \hline 
 1 & 4 & $3+4 x^2+x^4$  & $\left\{\sqrt{2},\frac{1}{\sqrt{2}},\sqrt{\frac{3}{2}}\right\}$  \\ \hline \hline
 2 & 4 & $1+4 x^2+x^4$  & $\left\{\sqrt{2},\sqrt{\frac{3}{2}},\frac{1}{\sqrt{2}}\right\}$  \\ \hline
 2 & 5 & $3+5 x^2+x^4$  & $\left\{\sqrt{\frac{5}{2}},\sqrt{\frac{13}{10}},\sqrt{\frac{6}{5}}\right\}$ \\ \hline 
 2 & 6 & $6+6 x^2+x^4$  & $\left\{\sqrt{3},1,\sqrt{2}\right\}$ \\ \hline \hline
 3 & 5 & $1+5 x^2+x^4$ & $\left\{\sqrt{\frac{5}{2}},\sqrt{\frac{21}{10}},\sqrt{\frac{2}{5}}\right\}$ \\ \hline 
 3 & 6 & $3+6 x^2+x^4$ & $\left\{\sqrt{3},\sqrt{2},1\right\}$  \\ \hline 
 3 & 7 & $6+7 x^2+x^4$  & $\left\{\sqrt{\frac{7}{2}},\frac{5}{\sqrt{14}},2 \sqrt{\frac{3}{7}}\right\} $ \\ \hline 
 3 & 8 & $10+8 x^2+x^4$  & $\left\{2,\sqrt{\frac{3}{2}},\sqrt{\frac{5}{2}}\right\} $ \\ \hline 
\end{tabular}
\caption{Results of the Schmeisser algorithm for the couplings in the inhomogeneous Ising chain $w_\ell=\sqrt{c_\ell}$ for various $M$ and $p$ values. Also shown is the form of the Q-polynomial \eqref{qpol}.}
\label{tab:1}
\end{table}

\begin{figure}[h!]
\centering
\includegraphics[width=7.5cm]{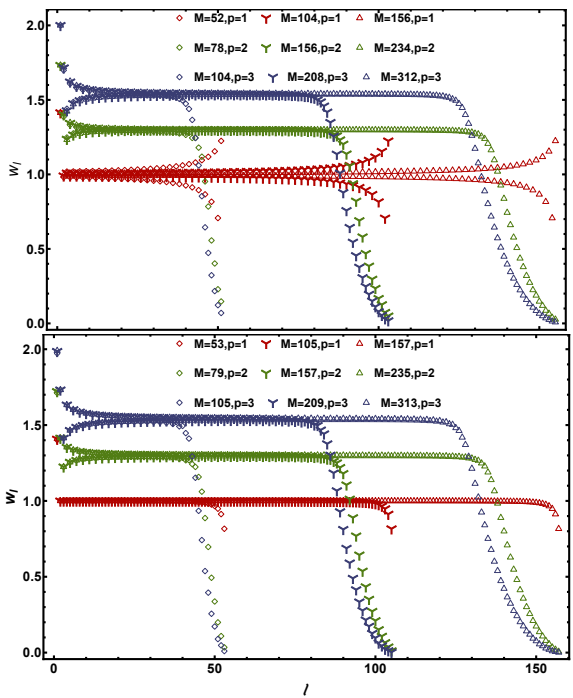}
\caption{Coefficients $w_\ell$ obtained from the Schmeisser algorithm for $p=1,2,3$ at the multicritical point $\lambda_{A,B,C,D}=1$. The values
of $M$ on the top and bottom panels are chosen such that $\mathfrak{p}_M=0$ and $\mathfrak{p}_M=1$,
respectively.}\label{fig:pic5}
\end{figure}

For larger values of $M$, we observe an interesting pattern of the $w$-couplings
associated with $\lambda_{A,B,C,...}=1$
for all values of $p$,
see Fig.~\ref{fig:pic5}. The couplings become
homogeneous in the bulk. Although not shown,
we verified that the behavior observed in Fig.~\ref{fig:pic5}
persists also for larger values of $d(M)\sim10^3$. We  also note that for the case $p=1$ the behavior near the edges is qualitatively different for odd and even $M$ values while there is no such qualitative difference for $p>1$. Furthermore, we want to remind the reader that the
case $p=1$, for which one of solutions of
(\ref{method}) is $w_\ell=\lambda_\ell\equiv 1$, is mapped to an {\it inhomogeneous}
quantum Ising chain $w_\ell\neq1$ if we use the Schmeisser algorithm.

\begin{figure}[h!]
\centering
\includegraphics[width=8.5cm]{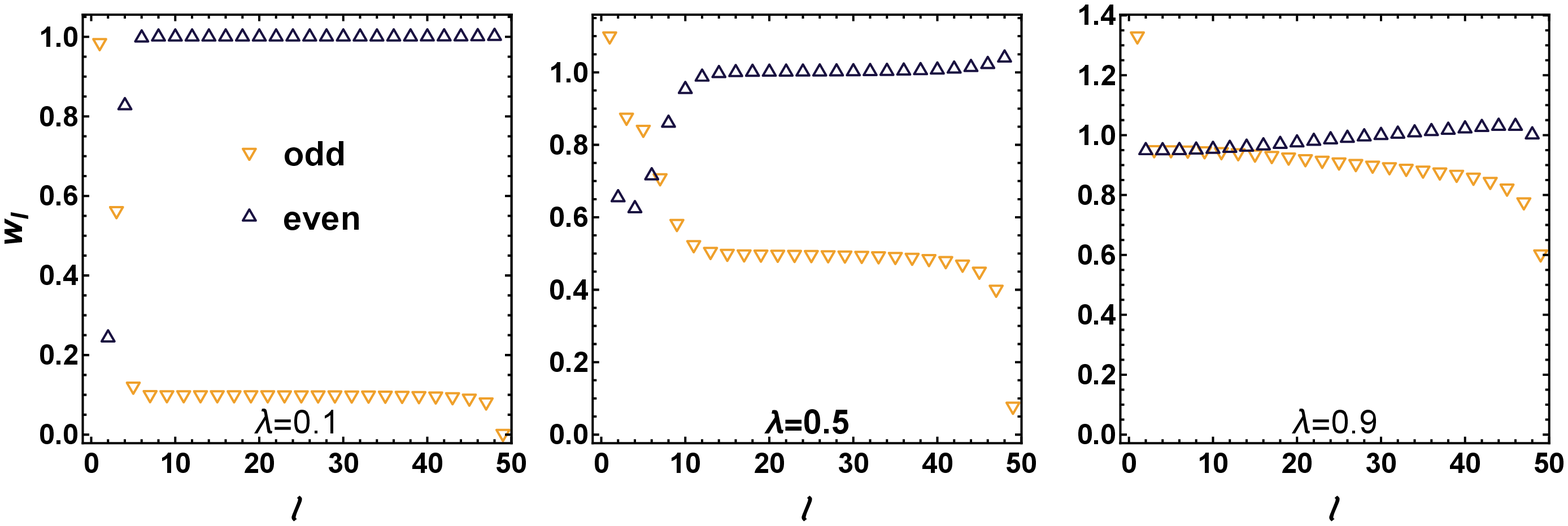}
\includegraphics[width=8.5cm]{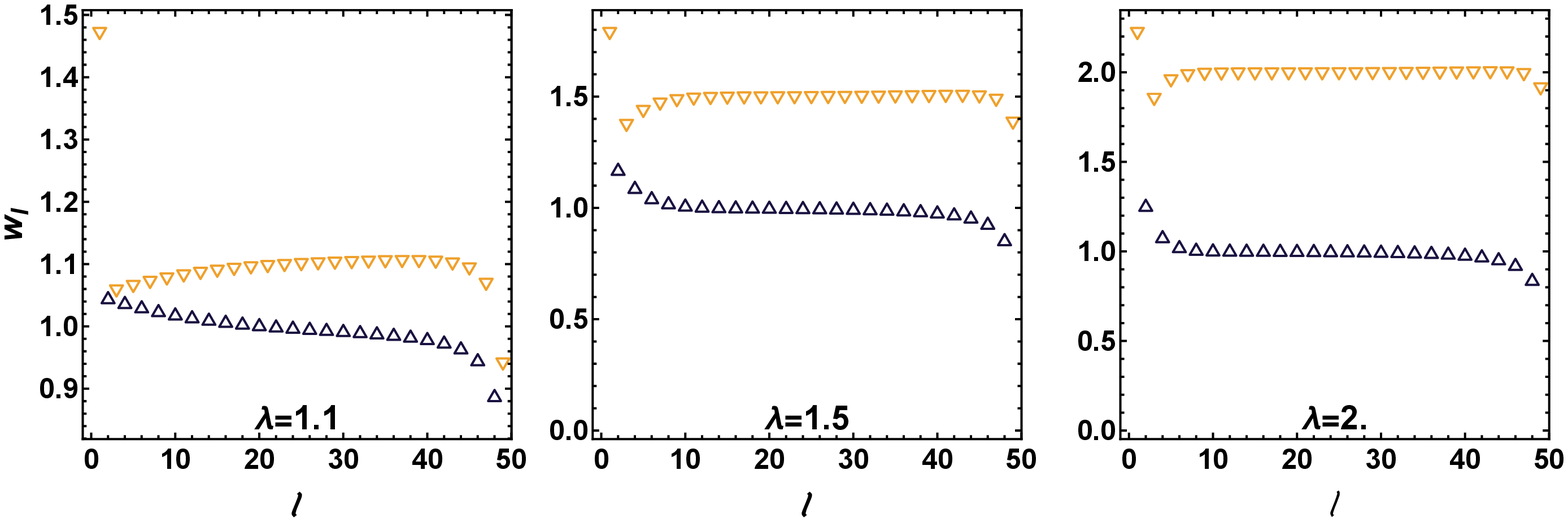}
    \caption{Coefficients $w_\ell$ for $p=1$ and $M=49$ for various
    values of the transverse field $\lambda\equiv \lambda_A/\lambda_B$ with different symbols for odd and even couplings.}\label{fig:pic6}
\end{figure}

\begin{figure}[h!]
\centering
\includegraphics[width=8.5cm]{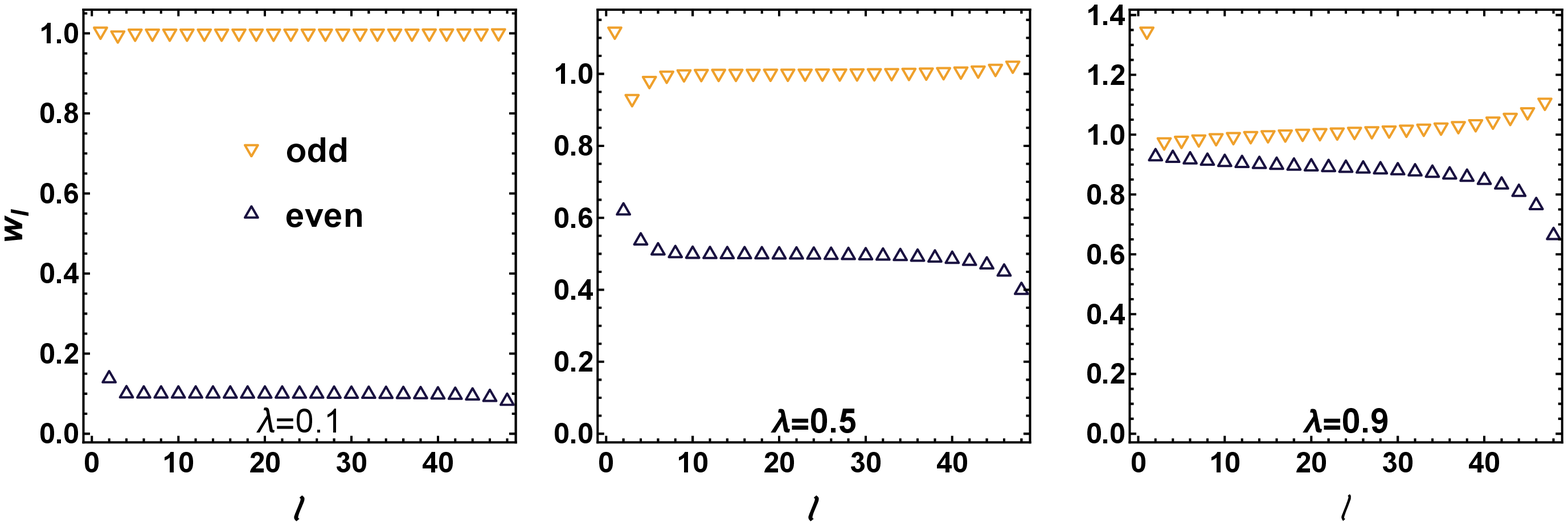}
\includegraphics[width=8.5cm]{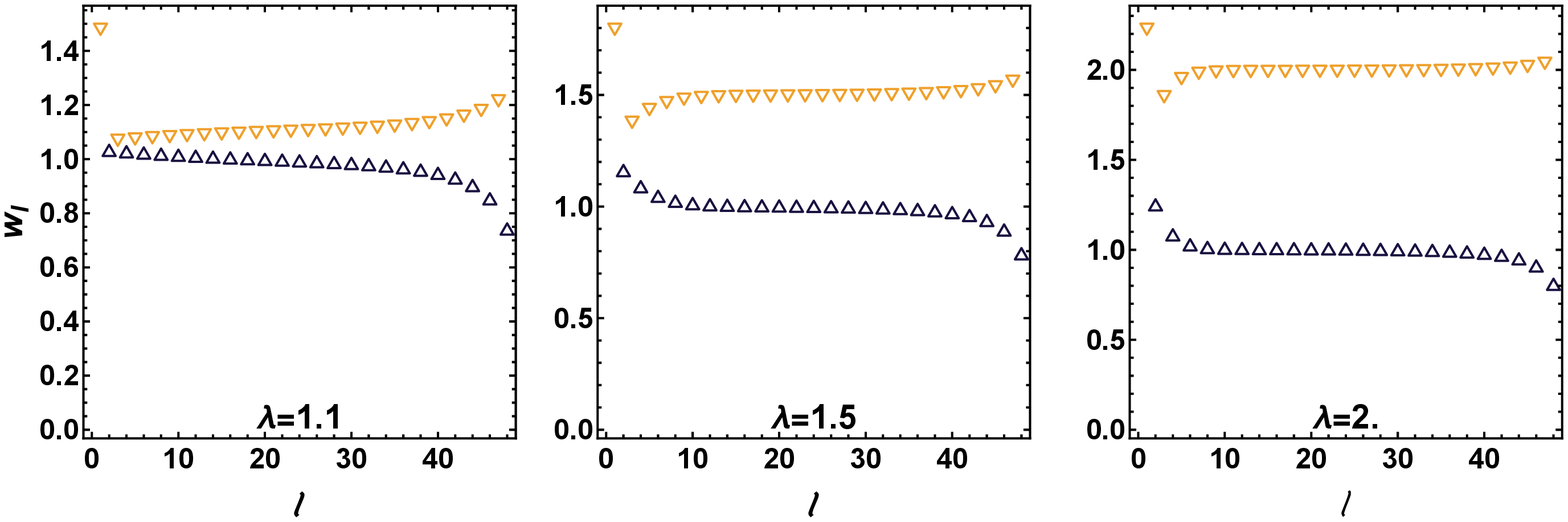}
    \caption{Coefficients $w_\ell$ for $p=1$ and $M=50$ for various
    values of the transverse field $\lambda\equiv \lambda_A/\lambda_B$ with different symbols for odd and even couplings.}\label{fig:pic7}
\end{figure}

We can also analyze the behavior of the coefficients $w_{\ell}$ when we move away
from criticality. In this case, the parity of the number 
of generators $M$ plays an important role. To start,
we consider the case $p=1$ (\ref{Hp1}) and set $\lambda\equiv\lambda_A/\lambda_B$ as the transverse field. For each value of $\lambda$ we generate the $Q$-polynomial (\ref{qpol}) and apply the
Schmeisser algorithm. The coefficients $w_\ell$ do depend on $\lambda$ and there is a qualitative change as we move away
from the critical value $\lambda=1$, see Fig.~\ref{fig:pic6} and Fig.~\ref{fig:pic7}.
For odd $M$, we have $w_{\text{even}}\approx 1$ and $w_{\text{odd}}\approx \lambda$ in the bulk.
Recall that the Ising chain (odd $M$) is ordered for $\lambda<1$ and 
disordered for $\lambda>1$. We then expect
the inhomogeneous model to be ordered if $w_{\text{even}}>w_{\text{odd}}$ and
disordered otherwise.
For even $M$, the algorithm gives 
$w_{\text{even}}<w_{\text{odd}}\approx 1$ if $\lambda<1$
and $1\approx w_{\text{even}}<w_{\text{odd}}$ for $\lambda>1$, see Fig.~\ref{fig:pic7}. This result
suggests disorder over all $\lambda\neq 1$ \cite{AHP21},
a fact which is consistent with the definition (\ref{qpol}) which excludes the exact zero mode.
Indeed, if one considers instead the polynomial (\ref{qpolt}) for even $M$, then the algorithm
returns $w_{\text{even}}>w_{\text{odd}}$ for all $\lambda\neq 1$, that is,
in this case we expect ordered phases for both $\lambda<1$ and $\lambda>1$. This is
consistent with the presence of an exact zero mode over all couplings $\lambda$, see Fig.~\ref{fig:pic2}.
In short, the coefficients given by the Schmeisser algorithm directly reflect the quantum phases and the quantum phase transition
at $\lambda=1$.

\begin{figure}[h!]
\centering
\includegraphics[width=8.5cm]{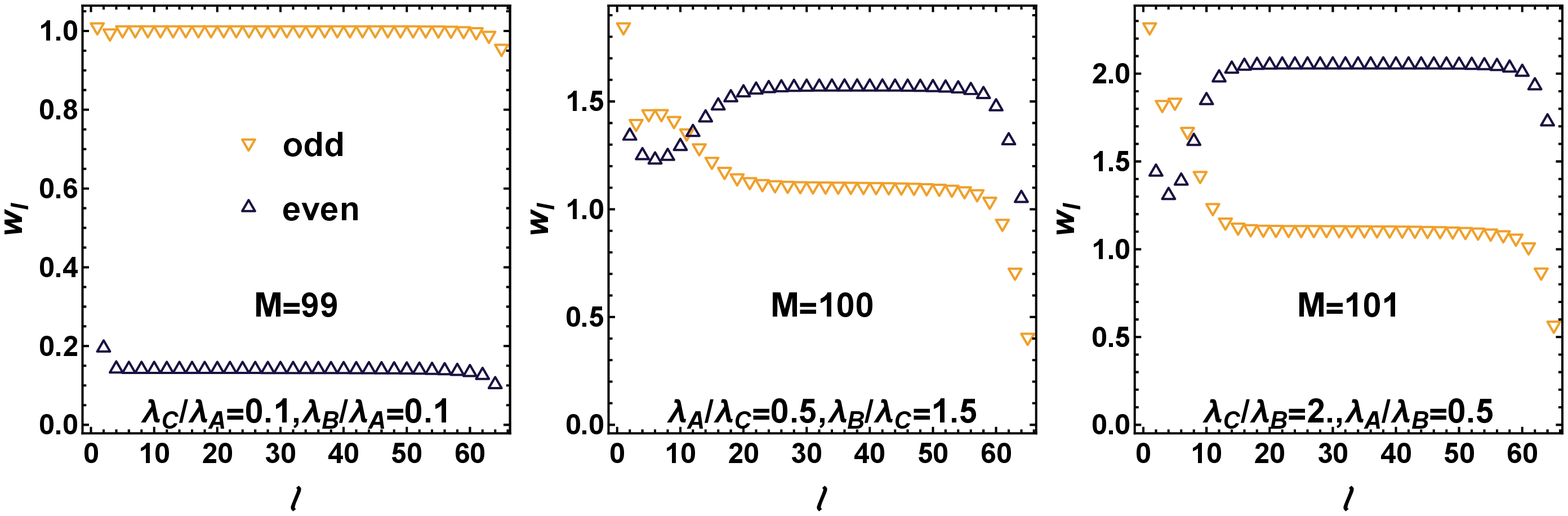}
    \caption{Coefficients $w_\ell$ for $p=2$ and different parities of $M$ and different couplings $\lambda_{A,B,C}$.}
    \label{fig:pic8}
\end{figure}

In the case $p=2$, the profile of the coefficients of the inhomogeneous Ising chain depends on the parity $\mathfrak{p}_M$. It is consistent
with the limiting case $p=1$ as well as the zero mode
regions in Fig.~\ref{fig:pic3} for the corresponding multispin chain. Wherever there is a zero mode,
we expect an ordered phase characterized by $w_{\text{even}}>w_{\text{odd}}$.
For the case $\mathfrak{p}_M=0$, see the first row of Fig.~\ref{fig:pic3},
all three gapped regions are expected to be disordered, and the algorithm
indeed gives $w_{\text{even}}<w_{\text{odd}}$ for the corresponding inhomogeneous Ising chain in all three regions. As an example,
the point
$\lambda_C/\lambda_A=0.1$ and $\lambda_B/\lambda_A=0.1$ is shown in the left panel
of Fig.~\ref{fig:pic8}.
If $\mathfrak{p}_M=1$, see the second row of Fig.~\ref{fig:pic3}, two out of
the three regions are expected to be ordered, while for $\mathfrak{p}_M=2$, see the third row of Fig.~\ref{fig:pic3}, only one region is expected to be ordered.
One can check that the algorithm returns the expected coefficient profiles in each of these cases. Along the critical lines, the profile of the coefficients are similar to those shown in Fig.~\ref{fig:pic5} for the $p=1$ case. Some representative cases for $\mathfrak{p}_M=1$ and $\mathfrak{p}_M=2$ are  also shown in Fig.~\ref{fig:pic8}.

\section{Correlations}\label{sec:corr}
To confirm the picture suggested by the ordering of the even and odd couplings described in the previous section, we now turn to a study of the spin-spin correlations 
of the effective inhomogeneous Ising chains. The computation
of correlation functions for quadratic Hamiltonians is possible at least numerically.
The building block is the two Majorana correlator
$\langle \psi_a\psi_b \rangle$. As we recall in App.~\ref{sec:standardd},
this correlator for the ground state is given in terms of eigenvectors $\vec{r},\, \vec{s}$ of the tridiagonal matrix (\ref{hopa}),
with eigenvalues $i\epsilon_k$ and $-i\epsilon_k$, respectively,
\be\label{exppsi}
\langle \psi_a\psi_b \rangle =2\sum_{k=1}^{\bar m} \frac{1}{N_k^2}r_{k,a}s_{k,b}\, .
\ee
This is valid for arbitrary couplings $w_\ell$. For the inhomogeneous models we have $\bar m=d(M)/2=\bar M$.
 
In this paper, we focus on the longitudinal spin-spin correlation, given
by \cite{SML64,P70}
\be\label{2p}
C_{a,b}^z\equiv\langle \sigma_a^z\sigma_b^z \rangle = (-1)^{b-a}
\det_{\substack{a\leq k'\leq b-1\\a+1\leq k''\leq b}}\left(i\langle \psi_{2k''-1}\psi_{2k'} \rangle\right)\,,\non\\
\ee
for any $a<b$. We first recall that for the standard homogeneous quantum Ising chain with transverse field $\lambda$, the correlation
(\ref{2p}) can be analytically evaluated in some cases.
For example, the nearest neighbor
correlation $C_{\ell,\ell+1}^z$ for the open
chain was
computed explicitly \cite{LHBZ19}. To investigate long-range correlations,
we can set $a=\ell$ and $b=\ell+R$ and then (\ref{2p}) can be exactly
evaluated for the homogeneous case in the thermodynamic limit \cite{P70}. This quantity is an order parameter for $R\to\infty$ and given by
\be\label{2pi}
\langle \sigma_\ell^z\sigma_{\ell+R}^z\rangle = \begin{cases}
    (1-\lambda^2)^{1/4}, & \lambda\leq 1,\\
    0, & \lambda>1.
  \end{cases}
\ee
For the inhomogeneous models constructed in Sec.~\ref{sec:inhomo},
obtaining analytical results for the correlations (\ref{2p}) in the thermodynamic
limit is an open problem. In the following, we present a numerical analysis.

\subsection{Site dependence}
We consider finite spin chains with open boundaries. Thus the expression
(\ref{2p}) is site dependent and boundary effects occur \cite{P19}. Since we are interested in the bulk behavior, we will mostly consider two-point correlation functions for sites which are far away from the boundaries.

\begin{figure}[h!]
\centering
\includegraphics[width=7.5cm]{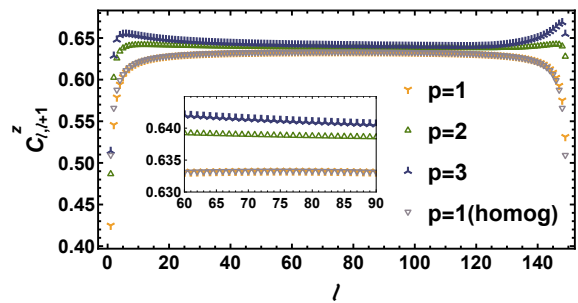}
    \caption{Nearest-neighbor correlations at the critical point $\lambda_{A,B,C,D}=1$ for $p=1,2,3$ and $M=299,448,597$, respectively. The standard homogeneous Ising chain
    at $\lambda=1$ is also shown.}
    \label{fig:pic9}
\end{figure}

\begin{figure}[h!]
\centering
\includegraphics[width=7.5cm]{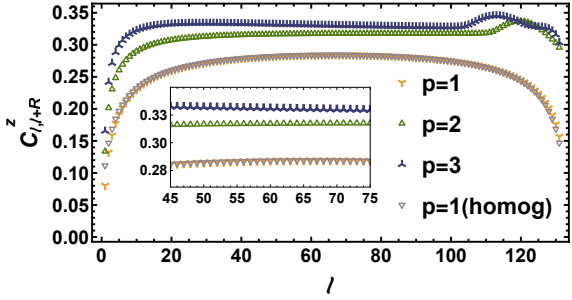}
    \caption{Long-range correlations at $\lambda_{A,B,C,D}=1$ for $p=1,2,3$ and $M=299,448,597$ with $R=\lfloor \bar M/8\rfloor =18$. The standard homogeneous Ising chain
    at $\lambda=1$ is also shown.}\label{fig:pic10}
\end{figure}

Let us first though consider the nearest-neighbor
correlation $C_{\ell,\ell+1}^z$ as a function of the site index $\ell$ for the inhomogeneous models
at the multicritical point $\lambda_{A,B,C,D}=1$ for $p=1,2,3$ with $M=299,449,599$ for $p=1,2,3$ respectively. For these values of $M$ we have $\mathfrak{p}_M=1$ and the number of sites is $\bar M=150$. As shown
in Fig.~\ref{fig:pic9}, we do observe a reasonable uniformity around
the center of the spin chain. Let us point out in particular, that the results for the inhomogeneous case $p=1$ are very similar to those for the standard homogeneous Ising chain. We verified that the form of the correlations in Fig.~\ref{fig:pic9} is qualitatively similar for different parities of $M$, except
for the case $p=1$ with even $M$. This is expected since the pattern of the couplings is different for this case as we see in
Fig.~\ref{fig:pic5}.
Similarly, we can consider long-range correlations, see Fig.~\ref{fig:pic10} for a distance between
the spins given by $R=\lfloor \bar M/8\rfloor =18$. For this type of correlation we observe stronger
boundary effects for $p=2,3$ as compared to $p=1$ due to the decay of the amplitude of the coefficients on the right
side of the chain, see Fig.~\ref{fig:pic5}.

\begin{figure}[h!]
\centering
\includegraphics[width=7cm]{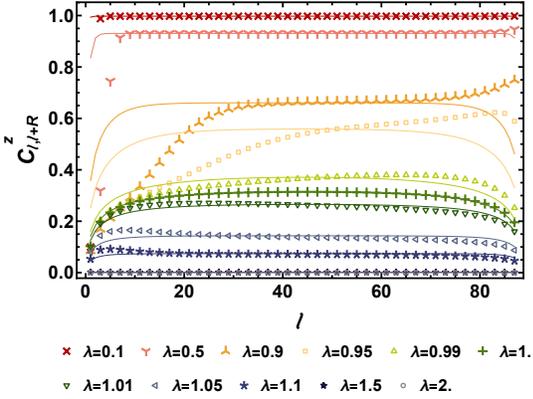}
    \caption{Long-range correlations as a function of the site position $\ell$, for $M=199$ ($\bar M=100$ sites), $p=1$, and $R=\lfloor \bar M/8 \rfloor=12$. The lines are the results for the standard homogeneous Ising chain. For better visualization, we show only half of the points for the inhomogeneous case.}\label{fig:pic12}
\end{figure}

We now also briefly consider the site dependence of the correlations (\ref{2p}) away from the multicritical point. For $p=1$, the long-range correlation $C^z_{\ell,\ell+R}$ with $R=12$ for various values
of the transverse field $\lambda\equiv \lambda_A/\lambda_B$ is shown in Fig.~\ref{fig:pic12}, along with the correlations computed for the standard homogeneous Ising chain. We note that the two results agree well inside the bulk except close to the critical point. Here we note that we keep the distance $R$ fixed while the correlation length in the inhomogeneous chain will change as a function of the site index $\ell$. We note, in particular, that the inhomogeneous chain does not have a reflection symmetry, see Fig.~\ref{fig:pic6}. The results are consistent with a transition from an ordered to a disordered phase.

\begin{figure}[h!]
\centering
\includegraphics[width=7cm]{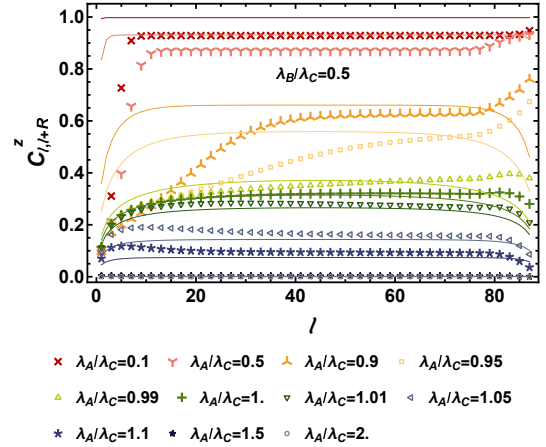}
    \caption{Long-range correlations as a function of the site position $\ell$, for $M=298$ ($\bar M=100$), $p=2$, $R=\lfloor \bar M/8 \rfloor=12$ and fixed $\lambda_B/\lambda_C=0.5$. The lines are the results for the standard homogeneous Ising chain with $M=199$. For better visualization, we show only half of the points for the inhomogeneous case.}\label{fig:pic13}
\end{figure}

\begin{figure}[h!]
\centering
\includegraphics[width=7cm]{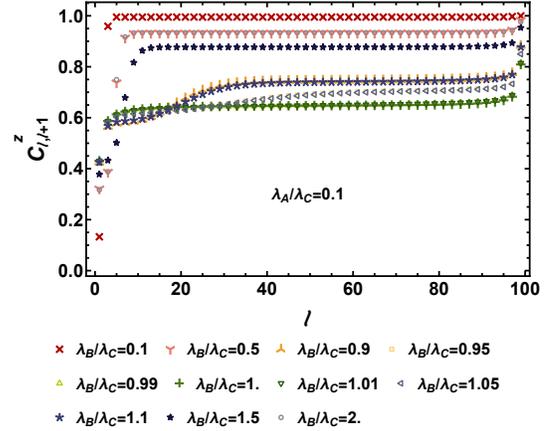}
    \caption{Long-range correlations as a function of the site position $\ell$, for $M=298$ ($\bar M=100$), $p=2$, $R=\lfloor \bar M/8 \rfloor=12$ and fixed $\lambda_A/\lambda_C=0.1$. For better visualization, we show only half of the points.}\label{fig:pic14}
\end{figure}

Next, we consider the case $p=2$. For simplicity, let us consider
$M$ such that $\mathfrak{p}_M=1$, and analyze lines
in the plane $(\lambda_A/\lambda_C,\lambda_B/\lambda_C)$. First, we fix $\lambda_B/\lambda_C <1$ and vary $\lambda_A/\lambda_C$.
In this case we are moving from a region with zero mode to a region without zero mode, see first panel in the second row of Fig.~\ref{fig:pic3}.
For a small value of $\lambda_B/\lambda_C$, we find that the system behaves similarly to the case $p=1$
with odd $M$. As $\lambda_B/\lambda_C$ increases, the correlations start to deviate from the $p=1$ result, see Fig.~\ref{fig:pic13}. The results are again consistent with the expected transition from an ordered to a disordered phase.
As a second example for $p=2$, let us now fix $\lambda_A/\lambda_C$ and vary
$\lambda_B/\lambda_C$ for the parity $\mathfrak{p}_M=1$. In this case, as we vary 
$\lambda_B/\lambda_C$, we are moving between two regions with zero modes (see first panel in the second row of Fig.~\ref{fig:pic3}). Therefore we expect the system to remain ordered which is consistent with the numerical results, see Fig.~\ref{fig:pic14}.
The results in this subsection confirm that: (i) despite
the inhomogeneity of the couplings, the correlations
are reasonably uniform around the center the chain, and (ii) that the presence or absence of zero modes and the ordering of the coefficients $w_\ell$ are indeed indicative on whether the system is ordered or disordered.

\subsection{Order parameter}
In this section, we focus on the correlation (\ref{2p}) deep inside
the chain. That is, we take $a=\lfloor \bar M/2 \rfloor$ and $b=\lfloor \bar M/2 \rfloor+R$
with $R=\lfloor \bar M/8 \rfloor$ and then analyze the dependence of this correlation on the coupling parameters. Note that this correlation will become an order parameter in the limit $\bar M\to\infty$.

\begin{figure}[]
\centering
\includegraphics[width=8.5cm]{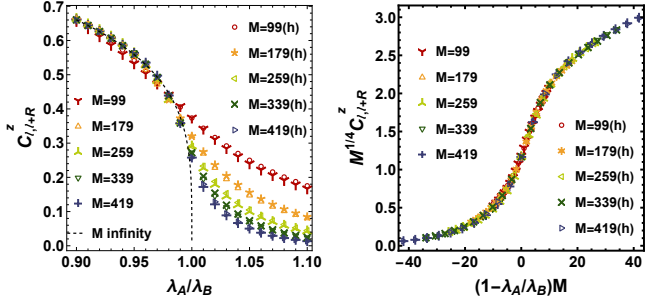}
\caption{The order parameter $C^z_{\ell,\ell+R}$ for the case $p=1$ for both the homogeneous (h)
and the inhomogeneous model. The values of $M$ correspond to lengths $\bar M=(M+1)/2$.
We set $\ell=\lfloor \bar M/2 \rfloor$ and $R=\lfloor \bar M/8 \rfloor$.}
\label{fig:pic15}
\end{figure}

We again start with the simplest case $p=1$ and set $\lambda\equiv\lambda_A/\lambda_B$.
We consider $M$ odd because for $M$ even (with the exact zero mode excluded)
the system is disordered for all $\lambda\neq 1$. Also, the odd $M$ case can be directly compared with
the standard homogeneous Ising case for which the behavior of the order parameter in the thermodynamic limit is known, see Eq.~(\ref{2pi}). In Fig.~\ref{fig:pic15}, left panel, we plot the long-range correlation $C^z_{\ell,\ell+R}$ for various
$M$ as a function of $\lambda$ for both the inhomogeneous
model and the standard homogeneous Ising chain. The correlations for the two systems become less and less distinguishable with increasing $M$ and approach the thermodynamic limit result (\ref{2pi}) shown as a dashed black line in Fig.~\ref{fig:pic15}. On the right panel, we show a scaling collapse of all the data for the homogeneous and inhomogenous models and various system lengths confirming the expected critical exponent $2\beta=1/4$ of the quantum Ising chain.

\begin{figure}[]
\centering
\includegraphics[width=8.5cm]{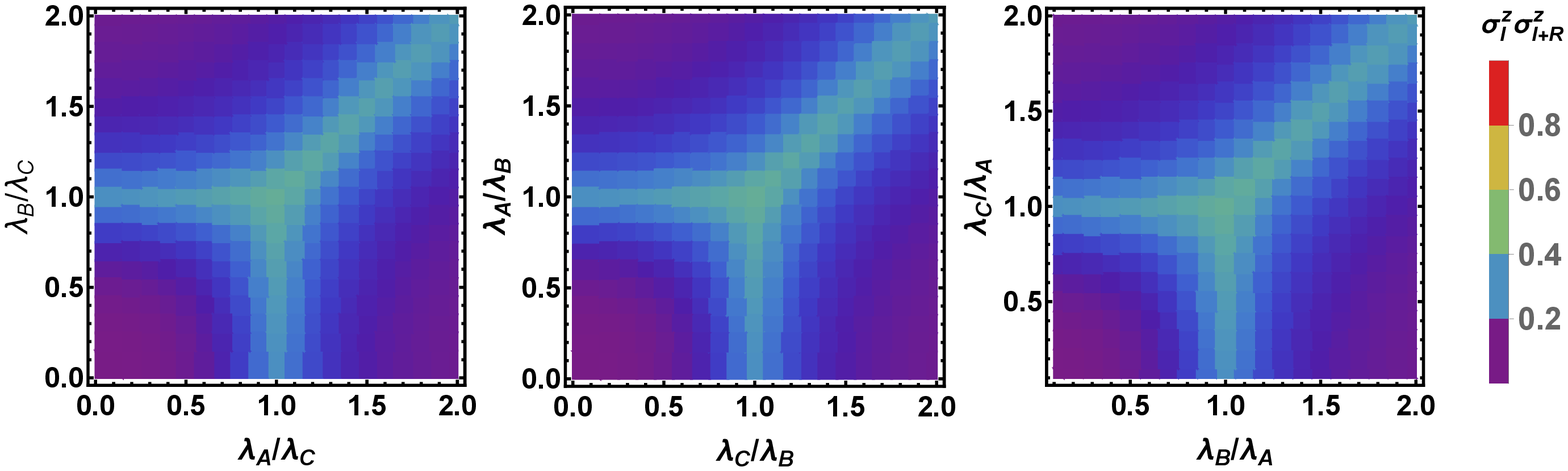}
(a) $M=99\,(\mathfrak{p}_M=0)$
\includegraphics[width=8.5cm]{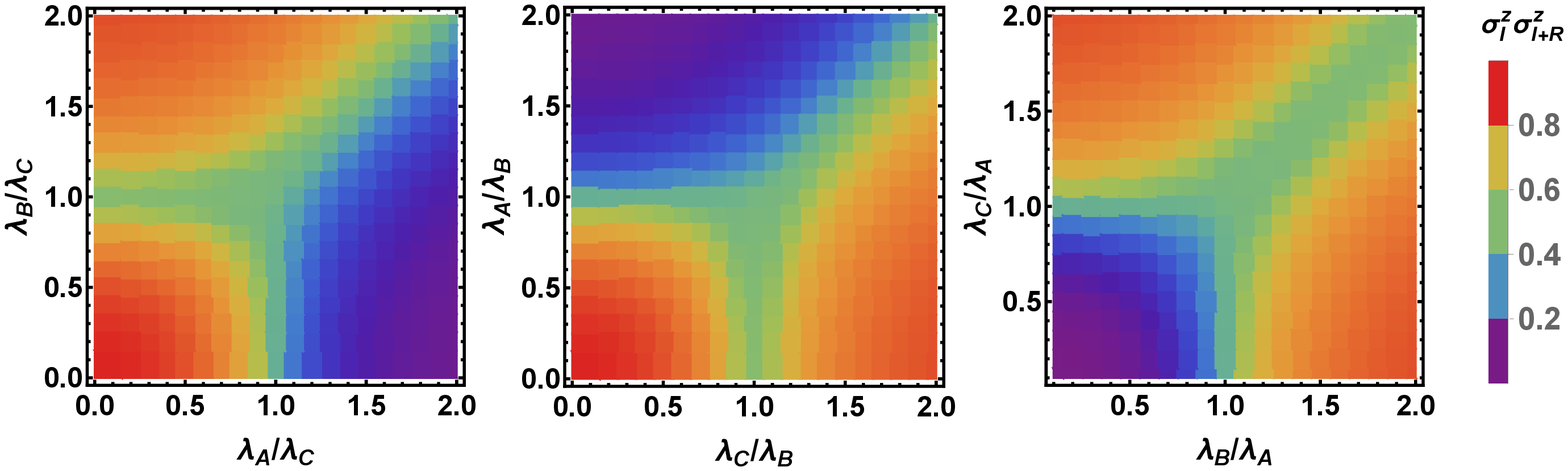}
(b) $M=100\,(\mathfrak{p}_M=1)$
\includegraphics[width=8.5cm]{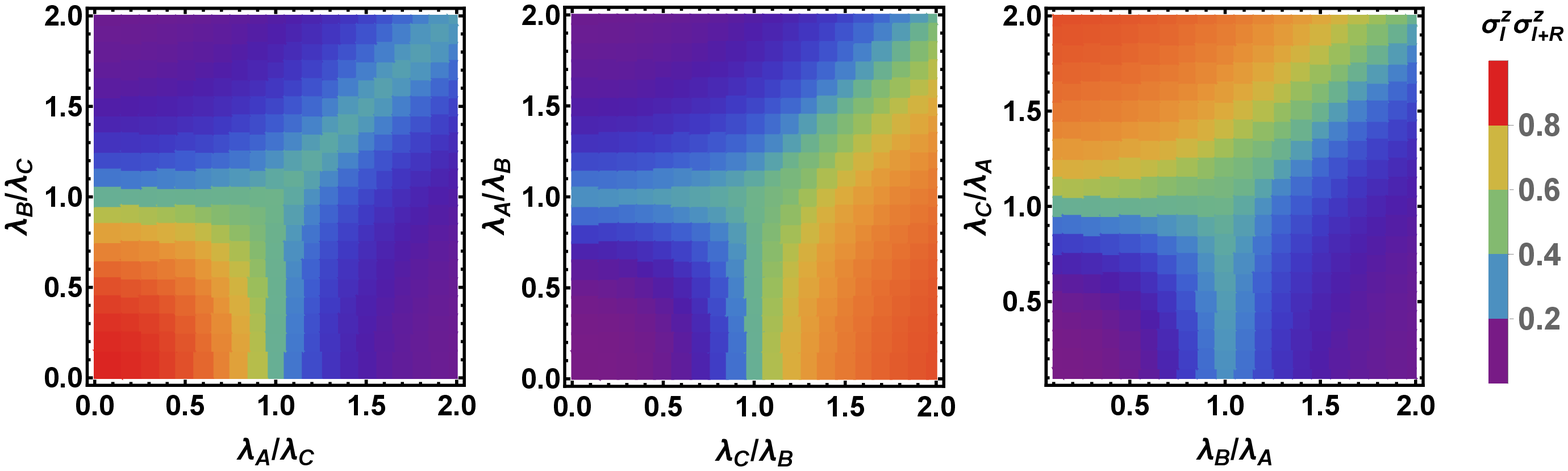}
(c) $M=101\,(\mathfrak{p}_M=2)$
\caption{Long-range correlator $C^z_{\ell,\ell+R}$ 
for $p=2$ and $M=99,100,101$ (corresponding to the parities $\mathfrak{p}_M=0,1,2$, respectively) as a function of
ratios of the coupling parameters $\{\lambda_A,\lambda_B,\lambda_C\}$. We set $\ell=\lfloor \bar M/2 \rfloor$ and $R=\lfloor \bar M/8 \rfloor=4$.}
\label{fig:pic16}
\end{figure}

We consider next the case $p=2$. First, we compute $C^z_{\ell,\ell+R}$ as a function of the
coupling parameters  $\{\lambda_A,\lambda_B,\lambda_C\}$ along different cuts,
see Fig.~\ref{fig:pic16}. As before, we consider the cases $M=99,100,101$ (length $\bar M=33,34,34$) which cover the three possible parities allowed in this case.
The phase diagram in Fig.~\ref{fig:pic16}
is in agreement with the zero mode pattern shown in Fig.~\ref{fig:pic3}. We used a grid with $\delta\lambda_j=0.1$
to produce the density plots. For $\mathfrak{p}_M=0$, the system is
disordered for all coupling parameters, except along the critical transition lines.
For the cases with $\mathfrak{p}_M\neq 0$ we observe
standard order-disorder transitions as in the homogeneous Ising chain but also order-order and disorder-disorder transitions.

\begin{figure}[]
\centering
\includegraphics[width=8.5cm]{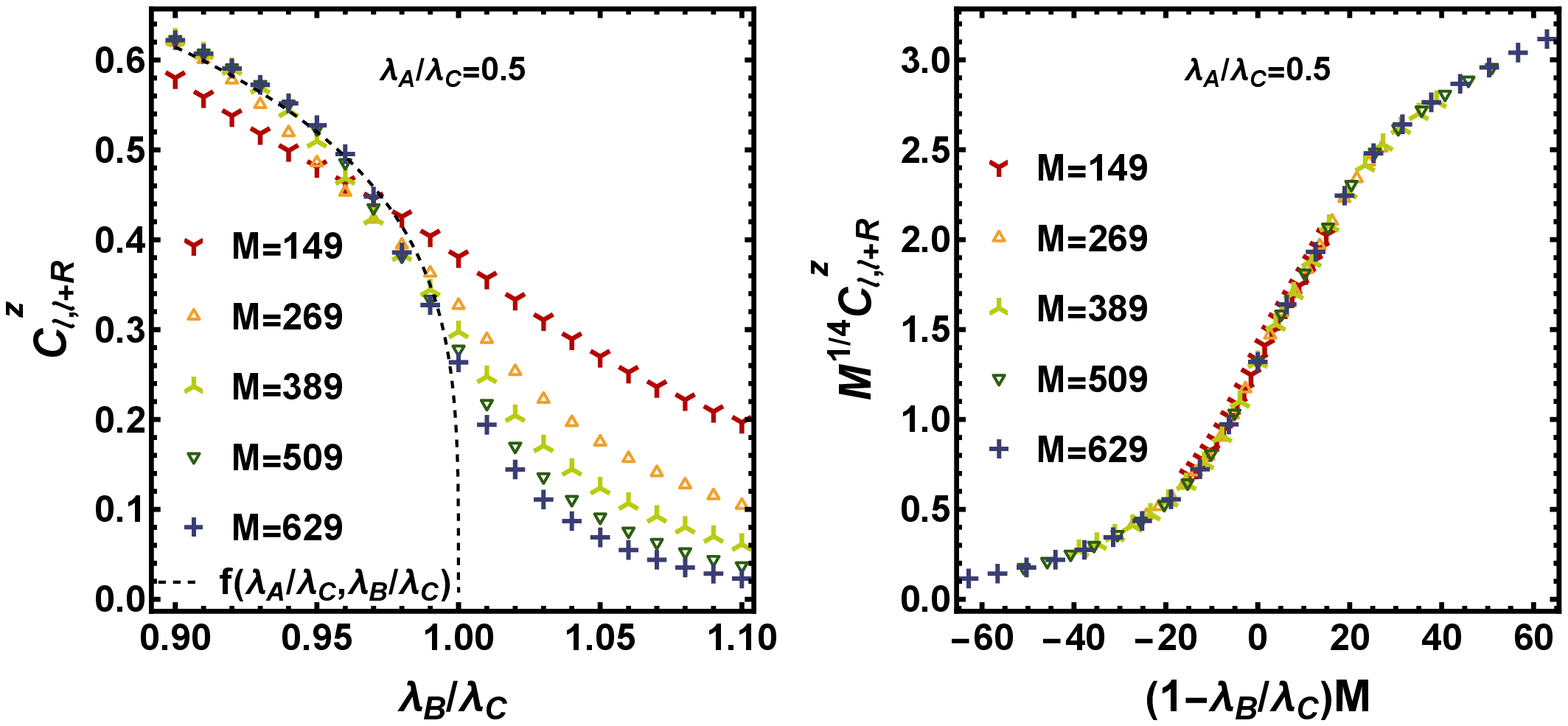}
\includegraphics[width=8.5cm]{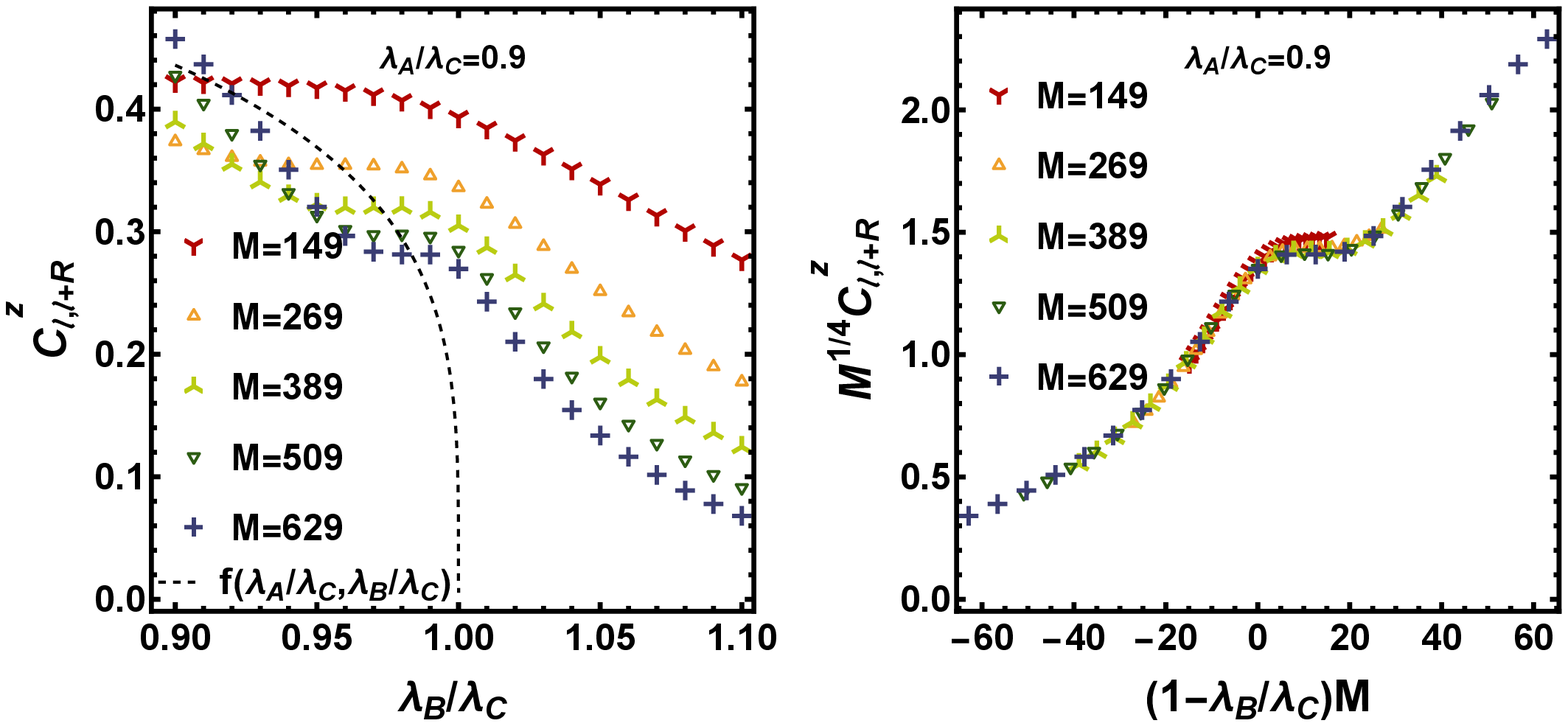}
\caption{$C^z_{\ell,\ell+R}$ as a function of $\{\lambda_A,\lambda_B,\lambda_C\}$ for $p=2$ and
various values of $M$ corresponding to $\bar M=34$. We set $\ell=\lfloor \bar M/2 \rfloor$ and $R=\lfloor \bar M/8 \rfloor$.}\label{fig:pic17}
\end{figure}

We now analyze some cuts in the coupling parameter space $\{\lambda_A,\lambda_B,\lambda_C\}$
for large values of $M$. As already discussed earlier, we can consider limiting cases where one of the split couplings
is small to understand the phase transitions. For instance, let us consider the case
where $\mathfrak{p}_M=2$, and let us fix $\lambda_A/\lambda_C$
while varying $\lambda_B/\lambda_C$, see the left panel in the last row of Fig.~\ref{fig:pic16}.
As $\lambda_A/\lambda_C\rightarrow0$, the system behaves similarly to the case $p=1$ with
odd $M$, that is, similarly to the quantum Ising chain depicted in Fig.~\ref{fig:pic15}.
As $\lambda_A/\lambda_C$ increases, the correlation is deformed. Remarkably, in the ordered
region $\lambda_A/\lambda_C<1,\,\lambda_B/\lambda_C<1$ but away from the critical line, the correlation is well described by
\be\label{ansa1}
\langle \sigma_\ell^z\sigma_{\ell+R}^z\rangle=f(\lambda_A/\lambda_C,\lambda_B/\lambda_C)
\ee
with
\begin{equation}
\label{fdef}
f(\lambda_A/\lambda_C,\lambda_B/\lambda_C) =\left(1-\frac{\lambda_A^2}{\lambda_C^2}\right)^{1/4}\left(1-\frac{\lambda_B^2}{\lambda_C^2}\right)^{1/4}.
\end{equation}
As an example, the point $\lambda_A/\lambda_C=0.5$ is shown in upper left panel of Fig.~\ref{fig:pic17}. However,
the order parameter profile deforms as the critical region $\lambda_A/\lambda_C\sim 1$
is approached, see the point $\lambda_A/\lambda_C=0.9$ in the bottom left panel of Fig.~\ref{fig:pic17}. Nevertheless, the curves in both cases show a scaling 
collapse with the same exponent $2\beta=1/4$. Therefore the more complicated profile close to the multicritical point does not indicate a change in the critical exponent but rather a scaling function which becomes more complex than \eqref{fdef}.

\begin{figure}[]
\centering
\includegraphics[width=8.5cm]{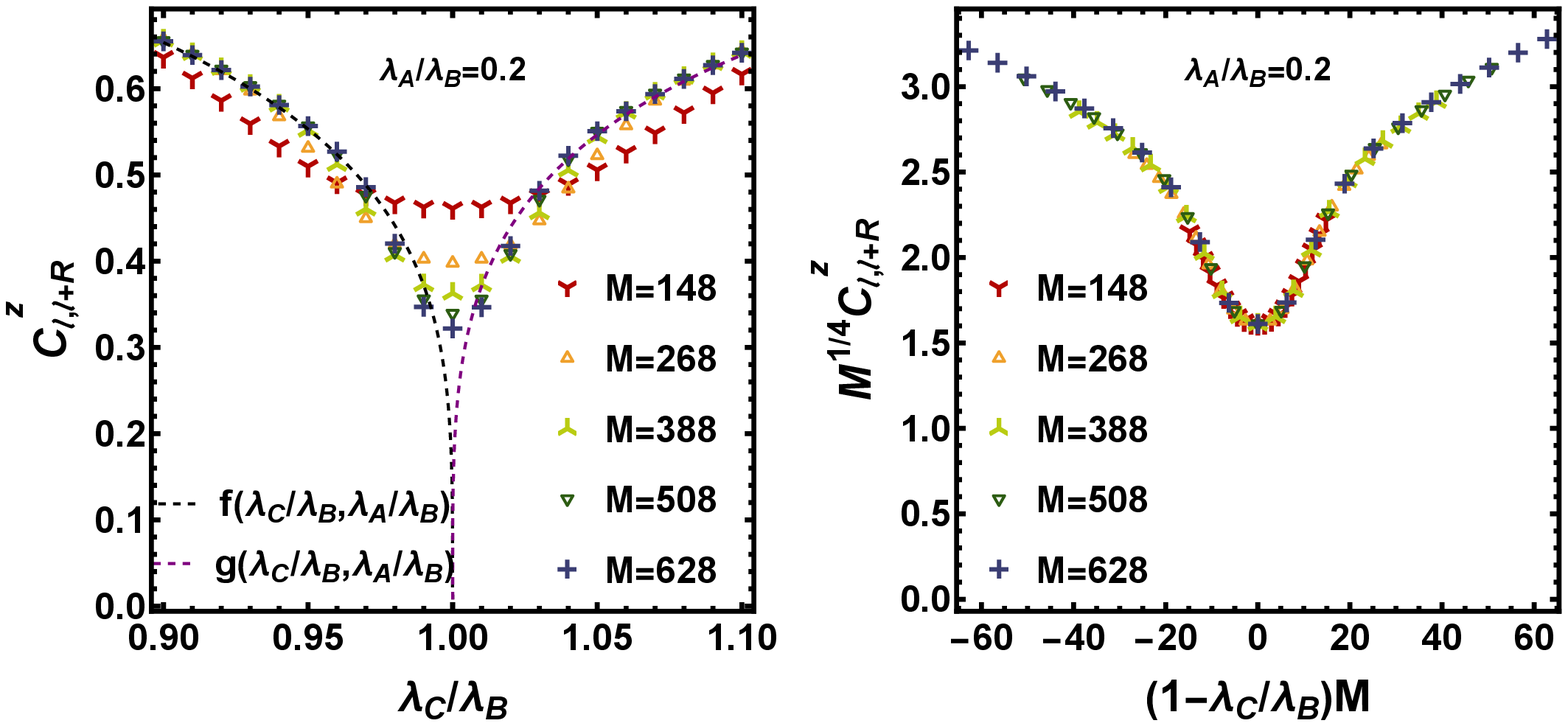}
\includegraphics[width=8.5cm]{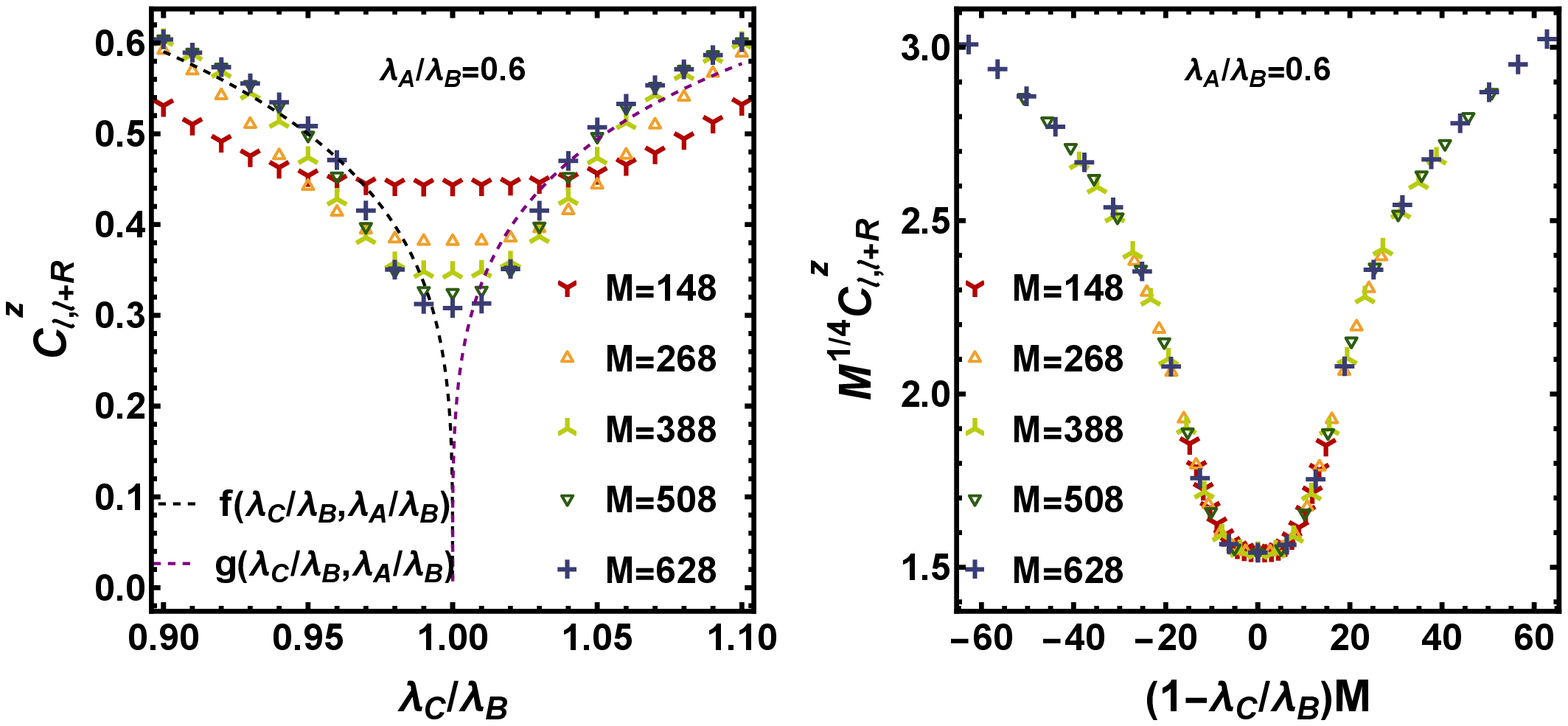}
\caption{$C^z_{\ell,\ell+R}$ as a function of $\{\lambda_A,\lambda_B,\lambda_C\}$ for $p=2$ and
various values of $M$ corresponding to $\bar M =34$. We set $\ell=\lfloor \bar M/2 \rfloor$ and $R=\lfloor \bar M/8 \rfloor$.}
\label{fig:pic18}
\end{figure}

Now, let us consider $\mathfrak{p}_M=1$ and fix $\lambda_A/\lambda_B$
while varying $\lambda_C/\lambda_B$, see the central panel in the middle row of Fig.~\ref{fig:pic16}.
For small $\lambda_A/\lambda_B$, we have a transition described by $p=1$ and even $M$, including
a zero mode over all $\lambda_C/\lambda_B$. In order words,
one has a transition of ordered-ordered type. This is related to the fact that
the polynomial $Q_M(x)$ (\ref{qpol}) for $p=2$ reduces to the polynomial $\tilde{Q}_M(x)$ (\ref{qpolt}) for
$p=1$ in the limit $\lambda_A/\lambda_B\rightarrow 0$. As $\lambda_A/\lambda_B$ increases, the order
parameter changes, as shown in Fig.~\ref{fig:pic18}. The following ansatz appears
to describe the order parameter well close to the axis $\lambda_A/\lambda_B\rightarrow0$
\begin{equation}
\langle \sigma_\ell^z\sigma_{\ell+R}^z\rangle = \begin{cases}
    f(\lambda_C/\lambda_B,\lambda_A/\lambda_B), & \lambda_C/\lambda_B< 1,\\
    g(\lambda_C/\lambda_B,\lambda_A/\lambda_B), & \lambda_C/\lambda_B> 1,
  \end{cases}
\end{equation}
where
\be
g(\lambda_C/\lambda_B,\lambda_A/\lambda_B)=\left(1-\frac{\lambda_A^2}{\lambda_B^2}\right)^{1/4}\left(1-\frac{\lambda_B^2}{\lambda_C^2}\right)^{1/4}\,.
\ee
and the function $f$ is given
by Eq.~(\ref{fdef}). In all cases we find a scaling collapse with the Ising critical exponent
$2\beta=1/4$.

A closed formula for the long-range correlation close to the critical lines and to the multicritical point is highly desirable, but is challenging to obtain. 
We 
can say, however, that 
a scaling collapse 
happens
in all cases 
with the same exponent 
$2\beta=1/4$. We have confirmed this scaling further by analyzing
the decay of $C^z_{\ell,\ell+R}$ as a function of $M$ at the multicritical point $\lambda_A=\lambda_B=\lambda_C=1$.
For $\mathfrak{p}_M=1,2$, we find that $C^z_{\ell,\ell+R}\sim 1/M^{1/4}$. The case $\mathfrak{p}_M=0$ is more challenging to analyze  because the multicritical point is surrounded by disordered phases in this case. 
Our numerical analysis indicate values of 
the exponents which deviate by about 10\% from the expected.
We 
believe that this is just a finite-size 
effect 
and the exponent $2\beta=1/4$ would be again obtained if 
larger chain lengths 
would be considered.

\section{Conclusion}\label{sec:conclu}
The motivation 
of the present paper is the understanding of the physics of general multispin free-fermionic systems that cannot be mapped onto
bilinear fermionic models. Such systems might allow to obtain analytical results, or at least highly accurate numerical results for long chains, for phases and phase transitions which are otherwise difficult to study. They can show, in particular, multicritical points with dynamical critical exponents $z>1$, thus putting them outside the realm of conformal field theories. However, while efficient methods to calculate the spectrum are known, the eigenstates have so far remained elusive,
although a formula in terms of the transfer matrix is
known \cite{F19}. Numerical evaluations
are also challenging due to the large global degeneracy of the spectrum, which is in general hard to lift.
Furthermore, Wick's theorem cannot be applied. The calculation of form factors and correlation functions thus remains a major challenge.

Here we have argued that one way to make progress is to construct inhomogeneous Ising chain analogues of these chains with multispin interactions. The inhomogeneous Ising chains can be constructed in such way that they have exactly the same spectrum as the multispin chains, with the advantage of avoiding the
high degeneracy. However, the eigenstates will 
be different in general. By studying the zero modes and long-range correlation functions in these inhomogeneous chains we have shown---mostly for the $p=2$ case (Fendley model)---that they have the same phase diagram including a multicritical point and thus indeed offer new insights into the physics of the multichain spin chains. In particular, the calculation of two-point spin correlation functions revealed that the transitions between the different phases, ordered and disordered, are all of Ising type with critical exponent $2\beta=1/4$. 

For the future, it would be interesting to investigate dynamical correlation functions, in particular at the multicritical points with $z>1$, and to extend these considerations to 
the case of free parafermionic chains.

\section*{Acknowledgments}
We thank Jos\'e Hoyos for discussions.
The work of FCA was supported in part
by the Brazilian agencies FAPESP and CNPq. RAP and JS acknowledge support by the
German Research Council (DFG) via the Research Unit FOR
2316. JS also acknowledges support by the National Science and Engineering Resource Council (NSERC) of Canada via the Discovery Grant program. 

\appendix

\section{Inhomogeneous Majorana chains}\label{sec:standardd}

In this section, we recall the solution of the quantum Ising chain in terms
of Majoranas, as outlined in Ref.~\cite{K01}, see also Ref.~\cite{LRG04}. The quantum Ising chain
can be written as (\ref{quadratic})
\be\label{H}
H=i\sum_{\ell=1}^m w_{\ell}\psi_{\ell+1}\psi_\ell
\ee
with $m=2L-1$. In this appendix, for generality, we keep $m$ as an arbitrary number. In this way, we can also consider the case of the Hamiltonian with an even number of generators. The Hamiltonian (\ref{H}) can be written as,
\be\label{Hpvec}
H=-\frac{i}{2}{\vec\psi}^T T_m \vec\psi
\ee
where ${\vec\psi}^T = \left(\psi_1\,\,\psi_2\,\,\cdots \psi_{m+1}\right)$ and the $(m+1)\times(m+1)$ matrix $T_m$ is the tridiagonal hopping matrix given by (\ref{hopa}), 
$(T_m)_{ij} = w_i\delta_{j,i+1}-w_{i-1}\delta_{i,j+1}$.

Since $T_m$ is a real antisymmetric matrix, it can be block diagonalized
by an orthogonal matrix $O$, namely,
\be\label{Adecomp1}
T_m = O^T \left[\displaystyle{\bigoplus_{i=1}^{\frac{m+1}{2}}}\left(
\begin{array}{cc}
0 & -\epsilon_i\\
  \epsilon_i & 0
\end{array}
\right) \right]O\,
\ee
if $m$ is odd and,
\be\label{Adecomp2}
T_m = O^T \left[0\bigoplus_{i=1}^{\frac{m}{2}}\left(
\begin{array}{cc}
0 & -\epsilon_i\\
  \epsilon_i & 0
\end{array}
\right)\right] O\,
\ee
if $m$ is even. The quasi-energies $i\epsilon_k$ are the roots of the
characteristic polynomial of the hopping matrix $T_m$, $W_m(x)=\det_{m+1}(T_m-x)$,
that is, $W_m(i\epsilon_k)=W_m(-i\epsilon_k)=0$.
We consider the ordering
$\epsilon_1<\epsilon_2<\cdots < \epsilon_{\bar m}$. To build $O$,
let us suppose that $\vec{r}_k=\left(r_{k,1}~~r_{k,2}~~\cdots~~ r_{k,m+1}\right)^T$ is an eigenvector of $T_m$ with eigenvalue $i\epsilon_k$ while $\vec{s}_k=\left(s_{k,1}~~s_{k,2}~~\cdots~~ s_{k,m+1}\right)^T$
is an eigenvector of $T_m$ with eigenvalue $-i\epsilon_k$. Then, we obtain the following
difference equations,
\be\label{difeq}
&&w_nr_{k,n+1}-w_{n-1}r_{k,n-1}=i\epsilon r_{k,n}\,,\non\\
&&w_ns_{k,n+1}-w_{n-1}s_{k,n-1}=-i\epsilon s_{k,n}\,.
\ee
The difference equations are solved by,
\be
&&r_{k,j}=(-1)^{j}\left(\prod_{a=j}^mw_a\right) W_{j-2}(i\epsilon_k)\,,\non\\
&&s_{k,j}=(-1)^{j}\left(\prod_{a=j}^mw_a\right) W_{j-2}(-i\epsilon_k)\,,
\ee
for $j=1,\dots,m+1$.
Let $\vec{R}_k=\vec{r}_k/N_k$ and $\vec{S}_k=\vec{s}_k/N_k$ be the normalized
vectors with
\be
N_k^2=\sum_{j=1}^{m+1} \left(\prod_{a=j}^mw_a\right)^2 W_{j-2}(i\epsilon_k)W_{j-2}(-i\epsilon_k)\,.
\ee
Now, let 
\be
\vec{o}_{2k-1}=\frac{1}{\sqrt{2}}\left(\vec{R}_k+\vec{S}_k\right),~ \vec{o}_{2k}=\frac{i}{\sqrt{2}}\left(\vec{R}_k-\vec{S}_k\right),
\ee
for $k=1,\dots,m+1$. For odd $m$, the matrix $O$ is the matrix with rows $\vec{o}_j^T$
with $j=1,\dots,m+1$. For even $m$, we have $\epsilon_1=0$ and $\vec{R}_1=\vec{S}_1$, such that $\vec{o}_1=2\vec{R}_1/\sqrt{2}$ and $\vec{o}_2=0$. Excluding the null row and renormalizing $\vec{o}_1\rightarrow\vec{o}_1=\vec{R}_1$ the matrix $O$ for even $m$ is the matrix with rows $\{\vec{o}_1^T,\vec{o}_3^T,\vec{o}_4^T,\dots,\vec{o}_{m+2}^T\}$.

Using $O$, we define
new Majorana operators
\be\label{newMaj}
\vec{\phi}=O\vec{\psi}\,,
\ee
such that the Hamiltonian (\ref{H}) is rewritten in decoupled modes,
\be
H=
i\sum_{k=1}^{\frac{m+1}{2}}\epsilon_k \phi_{2k-1}\phi_{2k}\,,
\ee
for odd $m$ and
\be
H=
i\sum_{k=2}^{\frac{m}{2}+1}\epsilon_k \phi_{2k-2}\phi_{2k-1}\,,
\ee
for even $m$.

Let us fix now, for simplicity, $\bar m=\frac{m+1}{2}$ for odd $m$.
In terms of complex fermions,
\be\label{phiPsi}
&&\phi_{2k-1}=\Psi_k^\dagger+\Psi_k\,\quad
\phi_{2k}=i\left(\Psi_k^\dagger-\Psi_k\right)
\ee
we have, for odd $m$,
\be
H=\sum_{k=1}^{\bar m}\epsilon_k\left[\Psi_k^\dagger, \Psi_k\right]\,.
\ee
Note that,
\be
\left[H,\Psi_k\right]=2\epsilon_k \Psi_k\,.
\ee
The ground state is given by,
\be\label{ground}
\Psi_k |0\rangle = \langle0 |\Psi_k^\dagger=0 \,.
\ee

Let us express the physical modes in terms of $\{\Psi_k,\Psi_k^\dagger\}$. Equation (\ref{newMaj}) can be written
as
\be\label{psilinear}
\psi_k &=& \sum_{k'=1}^{2\bar m} O_{k',k}\phi_{k'}=\sum_{k'=1}^{\bar m}\mathcal{O}_{k',k}\Psi_{k'}^\dagger+\mathcal{O}_{k',k}^*\Psi_{k'}\non\\
 &=&\sum_{k'=1}^{\bar m}\left(O_{2k'-1,k}+iO_{2k',k}\right)\Psi_{k'}^\dagger\non\\
&+&\left(O_{2k'-1,k}-iO_{2k',k}\right)\Psi_{k'}\,.
\ee
In terms of $r_{k,j}$ and $s_{k,j}$, we have,
\be
&&O_{2k'-1,k}=\frac{1}{\sqrt{2}N_{k'}}\left(r_{k',k}+s_{k',k}\right)\,,\non\\&&
O_{2k',k}=\frac{i}{\sqrt{2}N_{k'}}\left(r_{k',k}-s_{k',k}\right)\,,
\ee
leading to
\be
\psi_k &=&\sqrt{2}\sum_{k'=1}^{\bar m}\frac{s_{k',k}}{N_{k'}}\Psi_{k'}^\dagger+\frac{r_{k',k}}{N_{k'}}\Psi_{k'}\,.
\ee
We can then compute, from (\ref{ground}) and (\ref{fermions}),
\be
\langle \psi_a\psi_b \rangle =2\sum_{k=1}^{\bar m} \frac{1}{N_k^2}r_{k,a}s_{k,b}\,,
\ee
and define the correlation matrix with elements,
\be
G_{ab}=\langle \psi_a\psi_b \rangle-\delta_{a,b}\,,
\ee
where $a,b=1,\dots,2L$.

Finally, we can also express $\{\Psi_k,\Psi_k^\dagger\}$ in terms of $\psi_\ell$. Using again Eq.~(\ref{newMaj}), we obtain
\be
&&\Psi_k = \sum_{\ell=1}^{\bar m} \alpha_{k,\ell}\psi_{2\ell-1}+i\beta_{k,\ell}\psi_{2\ell}\,,\non\\
\quad
&&\Psi_k^\dagger = \sum_{\ell=1}^{\bar m} \alpha_{k,\ell}\psi_{2\ell-1}-i\beta_{k,\ell}\psi_{2\ell}\,,
\ee
where
\be
&&\alpha_{k,\ell} = \frac{1}{\sqrt{2}N_k}(r_{k,2\ell-1}+s_{k,2\ell-1})\,,\non\\
&&\beta_{k,\ell} = \frac{i}{\sqrt{2}N_k}(r_{k,2\ell}-s_{k,2\ell})\,.
\ee


\begin{thebibliography}{38}
\expandafter\ifx\csname natexlab\endcsname\relax\def\natexlab#1{#1}\fi
\expandafter\ifx\csname bibnamefont\endcsname\relax
  \def\bibnamefont#1{#1}\fi
\expandafter\ifx\csname bibfnamefont\endcsname\relax
  \def\bibfnamefont#1{#1}\fi
\expandafter\ifx\csname citenamefont\endcsname\relax
  \def\citenamefont#1{#1}\fi
\expandafter\ifx\csname url\endcsname\relax
  \def\url#1{\texttt{#1}}\fi
\expandafter\ifx\csname urlprefix\endcsname\relax\def\urlprefix{URL }\fi
\providecommand{\bibinfo}[2]{#2}
\providecommand{\eprint}[2][]{\url{#2}}

\bibitem[{\citenamefont{Onsager}(1944)}]{O44}
\bibinfo{author}{\bibfnamefont{L.}~\bibnamefont{Onsager}},
  \bibinfo{journal}{Physical Review} \textbf{\bibinfo{volume}{65}},
  \bibinfo{pages}{117} (\bibinfo{year}{1944}),
  \urlprefix\url{https://doi.org/10.1103/physrev.65.117}.

\bibitem[{\citenamefont{Kaufman}(1949)}]{K64}
\bibinfo{author}{\bibfnamefont{B.}~\bibnamefont{Kaufman}},
  \bibinfo{journal}{Physical Review} \textbf{\bibinfo{volume}{76}},
  \bibinfo{pages}{1232} (\bibinfo{year}{1949}),
  \urlprefix\url{https://doi.org/10.1103/physrev.76.1232}.

\bibitem[{\citenamefont{SCHULTZ et~al.}(1964)\citenamefont{SCHULTZ, MATTIS, and
  LIEB}}]{SML64}
\bibinfo{author}{\bibfnamefont{T.~D.} \bibnamefont{SCHULTZ}},
  \bibinfo{author}{\bibfnamefont{D.~C.} \bibnamefont{MATTIS}},
  \bibnamefont{and} \bibinfo{author}{\bibfnamefont{E.~H.} \bibnamefont{LIEB}},
  \bibinfo{journal}{Reviews of Modern Physics} \textbf{\bibinfo{volume}{36}},
  \bibinfo{pages}{856} (\bibinfo{year}{1964}),
  \urlprefix\url{https://doi.org/10.1103/revmodphys.36.856}.

\bibitem[{\citenamefont{Pfeuty}(1970)}]{P70}
\bibinfo{author}{\bibfnamefont{P.}~\bibnamefont{Pfeuty}},
  \bibinfo{journal}{Annals of Physics} \textbf{\bibinfo{volume}{57}},
  \bibinfo{pages}{79} (\bibinfo{year}{1970}),
  \urlprefix\url{https://doi.org/10.1016/0003-4916(70)90270-8}.

\bibitem[{\citenamefont{Kitaev}(2001)}]{K01}
\bibinfo{author}{\bibfnamefont{A.~Y.} \bibnamefont{Kitaev}},
  \bibinfo{journal}{Physics-Uspekhi} \textbf{\bibinfo{volume}{44}},
  \bibinfo{pages}{131} (\bibinfo{year}{2001}),
  \urlprefix\url{https://doi.org/10.1070/1063-7869/44/10s/s29}.

\bibitem[{\citenamefont{Schnyder et~al.}(2008)\citenamefont{Schnyder, Ryu,
  Furusaki, and Ludwig}}]{Schnyder_2008}
\bibinfo{author}{\bibfnamefont{A.~P.} \bibnamefont{Schnyder}},
  \bibinfo{author}{\bibfnamefont{S.}~\bibnamefont{Ryu}},
  \bibinfo{author}{\bibfnamefont{A.}~\bibnamefont{Furusaki}}, \bibnamefont{and}
  \bibinfo{author}{\bibfnamefont{A.~W.~W.} \bibnamefont{Ludwig}},
  \bibinfo{journal}{Physical Review B} \textbf{\bibinfo{volume}{78}}
  (\bibinfo{year}{2008}),
  \urlprefix\url{https://doi.org/10.1103%2Fphysrevb.78.195125}.

\bibitem[{\citenamefont{Fendley}(2019)}]{F19}
\bibinfo{author}{\bibfnamefont{P.}~\bibnamefont{Fendley}},
  \bibinfo{journal}{Journal of Physics A: Mathematical and Theoretical}
  \textbf{\bibinfo{volume}{52}}, \bibinfo{pages}{335002}
  (\bibinfo{year}{2019}),
  \urlprefix\url{https://doi.org/10.1088/1751-8121/ab305d}.

\bibitem[{\citenamefont{Kopp and Chakravarty}(2005)}]{Kopp_2005}
\bibinfo{author}{\bibfnamefont{A.}~\bibnamefont{Kopp}} \bibnamefont{and}
  \bibinfo{author}{\bibfnamefont{S.}~\bibnamefont{Chakravarty}},
  \bibinfo{journal}{Nature Physics} \textbf{\bibinfo{volume}{1}},
  \bibinfo{pages}{53} (\bibinfo{year}{2005}),
  \urlprefix\url{https://doi.org/10.1038%2Fnphys105}.

\bibitem[{\citenamefont{Niu et~al.}(2012)\citenamefont{Niu, Chung, Hsu, Mandal,
  Raghu, and Chakravarty}}]{Niu_2012}
\bibinfo{author}{\bibfnamefont{Y.}~\bibnamefont{Niu}},
  \bibinfo{author}{\bibfnamefont{S.~B.} \bibnamefont{Chung}},
  \bibinfo{author}{\bibfnamefont{C.-H.} \bibnamefont{Hsu}},
  \bibinfo{author}{\bibfnamefont{I.}~\bibnamefont{Mandal}},
  \bibinfo{author}{\bibfnamefont{S.}~\bibnamefont{Raghu}}, \bibnamefont{and}
  \bibinfo{author}{\bibfnamefont{S.}~\bibnamefont{Chakravarty}},
  \bibinfo{journal}{Physical Review B} \textbf{\bibinfo{volume}{85}}
  (\bibinfo{year}{2012}),
  \urlprefix\url{https://doi.org/10.1103%2Fphysrevb.85.035110}.

\bibitem[{\citenamefont{Chapman and Flammia}(2020)}]{CF20}
\bibinfo{author}{\bibfnamefont{A.}~\bibnamefont{Chapman}} \bibnamefont{and}
  \bibinfo{author}{\bibfnamefont{S.~T.} \bibnamefont{Flammia}},
  \bibinfo{journal}{Quantum} \textbf{\bibinfo{volume}{4}}, \bibinfo{pages}{278}
  (\bibinfo{year}{2020}),
  \urlprefix\url{https://doi.org/10.22331/q-2020-06-04-278}.

\bibitem[{\citenamefont{Alcaraz and Pimenta}(2020{\natexlab{a}})}]{AP20a}
\bibinfo{author}{\bibfnamefont{F.~C.} \bibnamefont{Alcaraz}} \bibnamefont{and}
  \bibinfo{author}{\bibfnamefont{R.~A.} \bibnamefont{Pimenta}},
  \bibinfo{journal}{Physical Review B} \textbf{\bibinfo{volume}{102}}
  (\bibinfo{year}{2020}{\natexlab{a}}),
  \urlprefix\url{https://doi.org/10.1103/physrevb.102.121101}.

\bibitem[{\citenamefont{Alcaraz and Pimenta}(2020{\natexlab{b}})}]{AP20b}
\bibinfo{author}{\bibfnamefont{F.~C.} \bibnamefont{Alcaraz}} \bibnamefont{and}
  \bibinfo{author}{\bibfnamefont{R.~A.} \bibnamefont{Pimenta}},
  \bibinfo{journal}{Physical Review B} \textbf{\bibinfo{volume}{102}}
  (\bibinfo{year}{2020}{\natexlab{b}}),
  \urlprefix\url{https://doi.org/10.1103/physrevb.102.235170}.

\bibitem[{\citenamefont{Baxter}(1989)}]{B89}
\bibinfo{author}{\bibfnamefont{R.}~\bibnamefont{Baxter}},
  \bibinfo{journal}{Physics Letters A} \textbf{\bibinfo{volume}{140}},
  \bibinfo{pages}{155} (\bibinfo{year}{1989}),
  \urlprefix\url{https://doi.org/10.1016/0375-9601%2889%2990884-0}.

\bibitem[{\citenamefont{Fendley}(2014)}]{F13}
\bibinfo{author}{\bibfnamefont{P.}~\bibnamefont{Fendley}},
  \bibinfo{journal}{Journal of Physics A: Mathematical and Theoretical}
  \textbf{\bibinfo{volume}{47}}, \bibinfo{pages}{075001}
  (\bibinfo{year}{2014}),
  \urlprefix\url{https://doi.org/10.1088/1751-8113/47/7/075001}.

\bibitem[{\citenamefont{Elman et~al.}(2021)\citenamefont{Elman, Chapman, and
  Flammia}}]{ECF20}
\bibinfo{author}{\bibfnamefont{S.~J.} \bibnamefont{Elman}},
  \bibinfo{author}{\bibfnamefont{A.}~\bibnamefont{Chapman}}, \bibnamefont{and}
  \bibinfo{author}{\bibfnamefont{S.~T.} \bibnamefont{Flammia}},
  \bibinfo{journal}{Communications in Mathematical Physics}
  \textbf{\bibinfo{volume}{388}}, \bibinfo{pages}{969} (\bibinfo{year}{2021}),
  \urlprefix\url{https://doi.org/10.1007/s00220-021-04220-w}.

\bibitem[{\citenamefont{Alcaraz et~al.}(2021)\citenamefont{Alcaraz, Hoyos, and
  Pimenta}}]{AHP21}
\bibinfo{author}{\bibfnamefont{F.~C.} \bibnamefont{Alcaraz}},
  \bibinfo{author}{\bibfnamefont{J.~A.} \bibnamefont{Hoyos}}, \bibnamefont{and}
  \bibinfo{author}{\bibfnamefont{R.~A.} \bibnamefont{Pimenta}},
  \bibinfo{journal}{Physical Review B} \textbf{\bibinfo{volume}{104}}
  (\bibinfo{year}{2021}),
  \urlprefix\url{https://doi.org/10.1103/physrevb.104.174206}.

\bibitem[{\citenamefont{Alcaraz and Pimenta}(2021)}]{AP21}
\bibinfo{author}{\bibfnamefont{F.~C.} \bibnamefont{Alcaraz}} \bibnamefont{and}
  \bibinfo{author}{\bibfnamefont{R.~A.} \bibnamefont{Pimenta}},
  \bibinfo{journal}{Phys. Rev. E} \textbf{\bibinfo{volume}{104}},
  \bibinfo{pages}{054121} (\bibinfo{year}{2021}),
  \urlprefix\url{https://link.aps.org/doi/10.1103/PhysRevE.104.054121}.

\bibitem[{\citenamefont{Miao}(2022)}]{YM22}
\bibinfo{author}{\bibfnamefont{Y.}~\bibnamefont{Miao}},
  \bibinfo{journal}{{SciPost} Physics} \textbf{\bibinfo{volume}{13}}
  (\bibinfo{year}{2022}),
  \urlprefix\url{https://doi.org/10.21468/scipostphys.13.3.070}.

\bibitem[{\citenamefont{Stokman}(2019)}]{S20}
\bibinfo{author}{\bibfnamefont{J.~V.} \bibnamefont{Stokman}},
  \bibinfo{journal}{Algebras and Representation Theory}
  \textbf{\bibinfo{volume}{23}}, \bibinfo{pages}{1523} (\bibinfo{year}{2019}),
  \urlprefix\url{https://doi.org/10.1007/s10468-019-09903-6}.

\bibitem[{\citenamefont{Gombor and Pozsgay}(2021)}]{GP21}
\bibinfo{author}{\bibfnamefont{T.}~\bibnamefont{Gombor}} \bibnamefont{and}
  \bibinfo{author}{\bibfnamefont{B.}~\bibnamefont{Pozsgay}},
  \bibinfo{journal}{Physical Review E} \textbf{\bibinfo{volume}{104}}
  (\bibinfo{year}{2021}),
  \urlprefix\url{https://doi.org/10.1103/physreve.104.054123}.

\bibitem[{\citenamefont{Chen et~al.}(2017)\citenamefont{Chen, Fradkin, and
  Witczak-Krempa}}]{Chen_2017}
\bibinfo{author}{\bibfnamefont{X.}~\bibnamefont{Chen}},
  \bibinfo{author}{\bibfnamefont{E.}~\bibnamefont{Fradkin}}, \bibnamefont{and}
  \bibinfo{author}{\bibfnamefont{W.}~\bibnamefont{Witczak-Krempa}},
  \bibinfo{journal}{Journal of Physics A: Mathematical and Theoretical}
  \textbf{\bibinfo{volume}{50}}, \bibinfo{pages}{464002}
  (\bibinfo{year}{2017}),
  \urlprefix\url{https://doi.org/10.1088%2F1751-8121%2Faa8dbc}.

\bibitem[{\citenamefont{{Nandy} et~al.}(2022)\citenamefont{{Nandy},
  {Lenar{\v{c}}i{\v{c}}}, {Ilievski}, {Mierzejewski}, {Herbrych}, and
  {Prelov{\v{s}}ek}}}]{2022arXiv221117181N}
\bibinfo{author}{\bibfnamefont{S.}~\bibnamefont{{Nandy}}},
  \bibinfo{author}{\bibfnamefont{Z.}~\bibnamefont{{Lenar{\v{c}}i{\v{c}}}}},
  \bibinfo{author}{\bibfnamefont{E.}~\bibnamefont{{Ilievski}}},
  \bibinfo{author}{\bibfnamefont{M.}~\bibnamefont{{Mierzejewski}}},
  \bibinfo{author}{\bibfnamefont{J.}~\bibnamefont{{Herbrych}}},
  \bibnamefont{and}
  \bibinfo{author}{\bibfnamefont{P.}~\bibnamefont{{Prelov{\v{s}}ek}}},
  \bibinfo{journal}{arXiv e-prints} \bibinfo{eid}{arXiv:2211.17181}
  (\bibinfo{year}{2022}), \eprint{2211.17181}.

\bibitem[{\citenamefont{Fiedler}(1990)}]{F90}
\bibinfo{author}{\bibfnamefont{M.}~\bibnamefont{Fiedler}},
  \bibinfo{journal}{Linear Algebra and its Applications}
  \textbf{\bibinfo{volume}{141}}, \bibinfo{pages}{265} (\bibinfo{year}{1990}),
  \urlprefix\url{https://doi.org/10.1016/0024-3795%2890%2990323-5}.

\bibitem[{\citenamefont{Schmeisser}(1993)}]{S93}
\bibinfo{author}{\bibfnamefont{G.}~\bibnamefont{Schmeisser}},
  \bibinfo{journal}{Linear Algebra and its Applications}
  \textbf{\bibinfo{volume}{193}}, \bibinfo{pages}{11} (\bibinfo{year}{1993}),
  \urlprefix\url{https://doi.org/10.1016/0024-3795%2893%2990268-s}.

\bibitem[{\citenamefont{Vinet and Zhedanov}(2012)}]{VZ12}
\bibinfo{author}{\bibfnamefont{L.}~\bibnamefont{Vinet}} \bibnamefont{and}
  \bibinfo{author}{\bibfnamefont{A.}~\bibnamefont{Zhedanov}},
  \bibinfo{journal}{Physical Review A} \textbf{\bibinfo{volume}{85}}
  (\bibinfo{year}{2012}),
  \urlprefix\url{https://doi.org/10.1103/physreva.85.012323}.

\bibitem[{\citenamefont{Cramp{\'{e}} et~al.}(2019)\citenamefont{Cramp{\'{e}},
  Nepomechie, and Vinet}}]{CNV19}
\bibinfo{author}{\bibfnamefont{N.}~\bibnamefont{Cramp{\'{e}}}},
  \bibinfo{author}{\bibfnamefont{R.~I.} \bibnamefont{Nepomechie}},
  \bibnamefont{and} \bibinfo{author}{\bibfnamefont{L.}~\bibnamefont{Vinet}},
  \bibinfo{journal}{Journal of Statistical Mechanics: Theory and Experiment}
  \textbf{\bibinfo{volume}{2019}}, \bibinfo{pages}{093101}
  (\bibinfo{year}{2019}),
  \urlprefix\url{https://doi.org/10.1088/1742-5468/ab3787}.

\bibitem[{\citenamefont{Finkel and Gonz{\'{a}}lez-L{\'{o}}pez}(2021)}]{FGL21}
\bibinfo{author}{\bibfnamefont{F.}~\bibnamefont{Finkel}} \bibnamefont{and}
  \bibinfo{author}{\bibfnamefont{A.}~\bibnamefont{Gonz{\'{a}}lez-L{\'{o}}pez}},
  \bibinfo{journal}{Journal of High Energy Physics}
  \textbf{\bibinfo{volume}{2021}} (\bibinfo{year}{2021}),
  \urlprefix\url{https://doi.org/10.1007/jhep12%282021%29184}.

\bibitem[{\citenamefont{Bernard et~al.}(2022)\citenamefont{Bernard,
  Cramp{\'{e}}, Nepomechie, Parez, d{'}Andecy, and Vinet}}]{BCNPAV22}
\bibinfo{author}{\bibfnamefont{P.-A.} \bibnamefont{Bernard}},
  \bibinfo{author}{\bibfnamefont{N.}~\bibnamefont{Cramp{\'{e}}}},
  \bibinfo{author}{\bibfnamefont{R.~I.} \bibnamefont{Nepomechie}},
  \bibinfo{author}{\bibfnamefont{G.}~\bibnamefont{Parez}},
  \bibinfo{author}{\bibfnamefont{L.~P.} \bibnamefont{d{'}Andecy}},
  \bibnamefont{and} \bibinfo{author}{\bibfnamefont{L.}~\bibnamefont{Vinet}},
  \bibinfo{journal}{Nuclear Physics B} \textbf{\bibinfo{volume}{984}},
  \bibinfo{pages}{115975} (\bibinfo{year}{2022}),
  \urlprefix\url{https://doi.org/10.1016/j.nuclphysb.2022.115975}.

\bibitem[{\citenamefont{de~Buruaga et~al.}(2019)\citenamefont{de~Buruaga,
  Santalla, Rodr{\'{\i}}guez-Laguna, and Sierra}}]{BSRS}
\bibinfo{author}{\bibfnamefont{N.~S.~S.} \bibnamefont{de~Buruaga}},
  \bibinfo{author}{\bibfnamefont{S.~N.} \bibnamefont{Santalla}},
  \bibinfo{author}{\bibfnamefont{J.}~\bibnamefont{Rodr{\'{\i}}guez-Laguna}},
  \bibnamefont{and} \bibinfo{author}{\bibfnamefont{G.}~\bibnamefont{Sierra}},
  \bibinfo{journal}{Journal of Statistical Mechanics: Theory and Experiment}
  \textbf{\bibinfo{volume}{2019}}, \bibinfo{pages}{093102}
  (\bibinfo{year}{2019}),
  \urlprefix\url{https://doi.org/10.1088/1742-5468/ab3192}.

\bibitem[{\citenamefont{Li and Yang}(2022)}]{LY22}
\bibinfo{author}{\bibfnamefont{C.}~\bibnamefont{Li}} \bibnamefont{and}
  \bibinfo{author}{\bibfnamefont{F.}~\bibnamefont{Yang}},
  \bibinfo{journal}{Frontiers of Physics} \textbf{\bibinfo{volume}{18}}
  (\bibinfo{year}{2022}),
  \urlprefix\url{https://doi.org/10.1007/s11467-022-1226-6}.

\bibitem[{\citenamefont{Ouvry and Polychronakos}(2022)}]{OP21}
\bibinfo{author}{\bibfnamefont{S.}~\bibnamefont{Ouvry}} \bibnamefont{and}
  \bibinfo{author}{\bibfnamefont{A.~P.} \bibnamefont{Polychronakos}},
  \bibinfo{journal}{Journal of Physics A: Mathematical and Theoretical}
  \textbf{\bibinfo{volume}{55}}, \bibinfo{pages}{485005}
  (\bibinfo{year}{2022}),
  \urlprefix\url{https://doi.org/10.1088/1751-8121/aca573}.

\bibitem[{\citenamefont{Minami}(2016)}]{M16}
\bibinfo{author}{\bibfnamefont{K.}~\bibnamefont{Minami}},
  \bibinfo{journal}{Journal of the Physical Society of Japan}
  \textbf{\bibinfo{volume}{85}}, \bibinfo{pages}{024003}
  (\bibinfo{year}{2016}),
  \urlprefix\url{https://doi.org/10.7566/jpsj.85.024003}.

\bibitem[{\citenamefont{Minami}(2021)}]{M21}
\bibinfo{author}{\bibfnamefont{K.}~\bibnamefont{Minami}},
  \bibinfo{journal}{Nuclear Physics B} \textbf{\bibinfo{volume}{973}},
  \bibinfo{pages}{115599} (\bibinfo{year}{2021}),
  \urlprefix\url{https://doi.org/10.1016/j.nuclphysb.2021.115599}.

\bibitem[{\citenamefont{Sirker et~al.}(2014)\citenamefont{Sirker, Maiti,
  Konstantinidis, and Sedlmayr}}]{SMKS14}
\bibinfo{author}{\bibfnamefont{J.}~\bibnamefont{Sirker}},
  \bibinfo{author}{\bibfnamefont{M.}~\bibnamefont{Maiti}},
  \bibinfo{author}{\bibfnamefont{N.~P.} \bibnamefont{Konstantinidis}},
  \bibnamefont{and} \bibinfo{author}{\bibfnamefont{N.}~\bibnamefont{Sedlmayr}},
  \bibinfo{journal}{Journal of Statistical Mechanics: Theory and Experiment}
  \textbf{\bibinfo{volume}{2014}}, \bibinfo{pages}{P10032}
  (\bibinfo{year}{2014}),
  \urlprefix\url{https://doi.org/10.1088/1742-5468/2014/10/p10032}.

\bibitem[{\citenamefont{Francica et~al.}(2016)\citenamefont{Francica, Apollaro,
  Gullo, and Plastina}}]{FAGP16}
\bibinfo{author}{\bibfnamefont{G.}~\bibnamefont{Francica}},
  \bibinfo{author}{\bibfnamefont{T.~J.~G.} \bibnamefont{Apollaro}},
  \bibinfo{author}{\bibfnamefont{N.~L.} \bibnamefont{Gullo}}, \bibnamefont{and}
  \bibinfo{author}{\bibfnamefont{F.}~\bibnamefont{Plastina}},
  \bibinfo{journal}{Physical Review B} \textbf{\bibinfo{volume}{94}}
  (\bibinfo{year}{2016}),
  \urlprefix\url{https://doi.org/10.1103/physrevb.94.245103}.

\bibitem[{\citenamefont{Liu et~al.}(2019)\citenamefont{Liu, Henry, Batchelor,
  and Zhou}}]{LHBZ19}
\bibinfo{author}{\bibfnamefont{Z.-Z.} \bibnamefont{Liu}},
  \bibinfo{author}{\bibfnamefont{R.~A.} \bibnamefont{Henry}},
  \bibinfo{author}{\bibfnamefont{M.~T.} \bibnamefont{Batchelor}},
  \bibnamefont{and} \bibinfo{author}{\bibfnamefont{H.-Q.} \bibnamefont{Zhou}},
  \bibinfo{journal}{Journal of Statistical Mechanics: Theory and Experiment}
  \textbf{\bibinfo{volume}{2019}}, \bibinfo{pages}{124002}
  (\bibinfo{year}{2019}),
  \urlprefix\url{https://doi.org/10.1088/1742-5468/ab4fe1}.

\bibitem[{\citenamefont{Pro{\'{s}}niak}(2019)}]{P19}
\bibinfo{author}{\bibfnamefont{O.~A.} \bibnamefont{Pro{\'{s}}niak}},
  \bibinfo{journal}{Physica Scripta} \textbf{\bibinfo{volume}{94}},
  \bibinfo{pages}{085201} (\bibinfo{year}{2019}),
  \urlprefix\url{https://doi.org/10.1088/1402-4896/ab1492}.

\bibitem[{\citenamefont{Latorre et~al.}(2004)\citenamefont{Latorre, Rico, and
  Vidal}}]{LRG04}
\bibinfo{author}{\bibfnamefont{J.}~\bibnamefont{Latorre}},
  \bibinfo{author}{\bibfnamefont{E.}~\bibnamefont{Rico}}, \bibnamefont{and}
  \bibinfo{author}{\bibfnamefont{G.}~\bibnamefont{Vidal}},
  \bibinfo{journal}{Quantum Information and Computation}
  \textbf{\bibinfo{volume}{4}}, \bibinfo{pages}{48} (\bibinfo{year}{2004}),
  \urlprefix\url{https://doi.org/10.26421/qic4.1-4}.

\end{thebibliography}

\end{document}